\documentclass[sigconf]{acmart}
\usepackage{xcolor}
\usepackage[normalem]{ulem}
\usepackage{pdfpages}

\AtBeginDocument{%
  }

\copyrightyear{2023} 
\acmYear{2023} 
\setcopyright{rightsretained} 
\acmConference[FAccT '23]{2023 ACM Conference on Fairness, Accountability, and Transparency}{June 12--15, 2023}{Chicago, IL, USA}
\acmBooktitle{2023 ACM Conference on Fairness, Accountability, and Transparency (FAccT '23), June 12--15, 2023, Chicago, IL, USA}
\acmDOI{10.1145/3593013.3594049}
\acmISBN{979-8-4007-0192-4/23/06}

\usepackage{mdframed}
\usepackage[export]{adjustbox}
\usepackage{float}
\usepackage[caption = false]{subfig}
\usepackage{graphicx}

\begin{document}

\title{Augmented Datasheets for Speech Datasets and Ethical Decision-Making}

\author{Orestis Papakyriakopoulos}
\authornote{Both authors contributed equally.}
\email{orestis.papakyriakopoulos@sony.com}
\orcid{1234-5678-9012}
\affiliation{%
  \institution{Sony AI}
  \streetaddress{}
  \city{Zurich}
  \state{}
  \country{Switzerland}
  \postcode{}
}

\author{Anna Seo Gyeong Choi}
\email{sc2359@cornell.edu}
\authornotemark[1]
\affiliation{%
  \institution{Cornell University}
  \streetaddress{}
  \city{Ithaca}
  \state{New York}
  \country{USA}}

\author{Jerone Andrews}
\email{jerone.andrews@sony.com}
\affiliation{%
  \institution{Sony AI}
  \streetaddress{}
  \city{Tokyo}
  \state{}
  \country{Japan}
}

\author{Rebecca Bourke}
\email{rebecca.bourke@sony.com}
\affiliation{%
  \institution{Sony AI}
  \streetaddress{}
  \city{Tokyo}
  \state{}
  \country{Japan}
}

\author{William Thong}
\email{william.thong@sony.com}
\affiliation{%
  \institution{Sony AI}
  \streetaddress{}
  \city{Zurich}
  \state{}
  \country{Switzerland}}

\author{Dora Zhao}
\email{dora.zhao@sony.com}
\affiliation{%
  \institution{Sony AI}
  \streetaddress{}
  \city{New York}
  \state{New York}
  \country{USA}
  \postcode{}}

\author{Alice Xiang}
\authornote{Co\textendash principal investigator.}
\email{alice.xiang@sony.com}
\affiliation{%
  \institution{Sony AI}
  \streetaddress{}
  \city{Seattle}
  \state{Washington}
  \country{USA}
  \postcode{}}

\author{Allison Koenecke}
\authornotemark[2]
\email{koenecke@cornell.edu}
\affiliation{%
  \institution{Cornell University}
  \city{Ithaca}
  \state{New York}
  \country{USA}}

\renewcommand{\shortauthors}{Trovato et al.}

\begin{abstract}
Speech datasets are crucial for training Speech Language Technologies (SLT); however, the lack of diversity of the underlying training data can lead to serious limitations in building equitable and robust SLT products, especially along dimensions of language, accent, dialect, variety, and speech impairment---and the intersectionality of speech features with socioeconomic and demographic features. Furthermore, there is often a lack of oversight on the underlying training data---commonly built on massive web-crawling and/or publicly available speech---with regard to the ethics of such data collection. To encourage standardized documentation of such speech data components, we introduce an augmented datasheet for \emph{speech} datasets\footnote{
Augmented datasheet templates and examples are available at \url{https://github.com/SonyResearch/project_ethics_augmented_datasheets_for_speech_datasets} \label{githuburl}
}, which can be used in addition to ``Datasheets for Datasets''~\cite{gebru2021datasheets}.  
We then exemplify the importance of each question in our augmented datasheet based on in-depth literature reviews of speech data used in domains such as machine learning, linguistics, and health. Finally, we encourage practitioners---ranging from dataset creators to researchers---to use our augmented datasheet to better define the scope, properties, and limits of speech datasets, while also encouraging consideration of data-subject protection and user community empowerment. Ethical dataset creation is not a one-size-fits-all process, but dataset creators can use our augmented datasheet to reflexively consider the social context of related SLT applications and data sources in order to foster more inclusive SLT products downstream.

\end{abstract}

\maketitle

\section{Introduction}

The ubiquity of Speech Language Technologies (SLT) in everyday life raises serious questions about the disparate harms these technologies can have on different populations. Use cases such as automated speech recognition (ASR), speaker recognition, speech synthesis, speech quality assessment, speech enhancement and denoising are now integrated in smartphones, cameras and virtual-assistants~\cite{wagner2019speech}, and applied in domains such as customer service~\cite{mckinsey2022}, finance~\cite{meng2004isis,gerz2021multilingual,pal2018pannomullokathan}, navigation \cite{dubey2022deep,yang2019atcspeech,yun2018automatic}, health~\cite{markoff2019}, education~\cite{yeung2019frequency,niknazar2021voice,papi2021mixtures,nishida2014promoting,myers2021}, and law~\cite{jaafari2019}.
The downstream impacts of biased SLT can be severe. Individuals who do not speak ``standard '' varieties of a language~\cite{Milroy2012} may be disproportionately unlikely to be hired given speech-based hiring screening software~\cite{Zuloaga2021}. Doctors increasingly use SLT to efficiently take patient notes, which could result in serious health harms if transcribed incorrectly~\cite{markoff2019}. Moreover, SLT technology has been developed to surveil the phone calls of incarcerated individuals; the resulting transcriptions---likely disproportionately inaccurate for Black individuals~\cite{Koenecke2020}---can result in differential treatment~\cite{Sherfinski2021}.

Speech technologies---as with any machine learning applications---are prone to bias; these biases often stem from nonrepresentative or inaccurate training data~\cite{Kim_2019_CVPR,Koh2021,Mitchell2021,Koenecke2020,buolamwini18a}. There is a mismatch between the ``world as it is,'' and the ``world according to data'' \cite{Mitchell2021}; there are further mismatches between the data used to train a machine learning model, and the data testing the model in practice \cite{Koh2021}. Generating robust SLT is an especially complex task given the diversity of human speech, comprising of different languages, accents, dialects, varieties, and speech impediments. An additional layer of complexity arises from the collection of speech data used to train SLT applications: there is high variability across necessary tasks, from noise in a recording environment to transcription of language and acoustic features. Finally, as with any dataset, it is imperative to center ethical dataset creation and usage---regarding the privacy, respect, and protection of data subjects, interviewers, and transcription annotators.

Given the complexity of creating, documenting, and using speech datasets, we propose ``augmented datasheets for speech datasets,'' inspired by the original ``datasheets for datasets'' \cite{gebru2021datasheets}. Datasheet usage can make dataset creation more transparent and accessible, while also assisting dataset users---such as SLT practitioners and researchers---to select appropriate datasets for their objectives. Our specific contributions are as follows:
first, we generate a set of speech data-specific questions for practitioners to answer when creating or using speech data (Section \ref{sec:examples}, bolded). Second, we substantiate why each question is important via examples of related papers and datasets (Section \ref{sec:examples}, plain text). Third, we make a call to action---for all practitioners using speech data, whether dataset creators or users---to use speech-specific datasheets in a collaborative effort to ensure transparency regarding speech data ethics and diversity (Section \ref{sec:discussion}). Fourth, we provide speech datasheet templates---in both .tex and .docx---both empty and with worked examples, for practitioner use (available on GitHub, Footnote \ref{githuburl}).

The questions comprising our ``augmented datasheet'' for speech datasets are formulated by the authors' positionalities as SLT practitioners, linguists,  machine learning researchers, algorithmic fairness researchers, and lawyers; we also draw upon an in-depth literature review revealing data-centric best practices and issues in SLT (Appendix \ref{litreview}). We believe our augmented datasheet template can guide researches in ethical speech dataset design and usage, and have released our template for immediate public use. However, we emphasize that our datasheet template should not replace the dire need for in-depth conversations on speech dataset ethics and diversity among datasheet users.

\subsection{Related Work \& Existing Datasheets}

Our study builds on prior work scrutinizing the design \cite{hutiri2022design,leclair2019recommendations,villazon2011methodological, andrews2023ethical,conspeech}, documentation \cite{gebru2021datasheets,srinivasan2021artsheets,pushkarna2022data,costa2020mt,rostamzadeh2022healthsheet,butcher2021causal}, and analytical evaluation \cite{mahajan2021need,fabris2022tackling,miceli2021documenting,schramowski2022can} of datasets.
For example, Hanley et al. \cite{hanley2020ethical} identified four aspects of human-centric dataset development that result in ethical concern, namely purpose (e.g., moral legitimacy), creation (e.g., data sourcing and cleaning), composition (e.g., data instances, metadata), and distribution (e.g., terms of use). Similarly, Mahajan et al. \cite{mahajan2021need} conducted a systematic review of over 300 research papers related to spoken or written multi-party dialogue, highlighting the need for diverse representation of human participants, privacy protection of sensitive information, and preregistration of proposed data collection processes to avoid purpose creep. A core cause of ethical concerns has been identified: opaque, sparse, and non-standardized documentation practices \cite{fabris2022tackling}.

Following common practices in the electronics industry, Gebru et al. \cite{gebru2021datasheets} proposed datasheets to document the key stages of a dataset’s lifecycle. Such documentation can help to address the reproducibility crisis by making ``research validity and integrity'' \cite{fabris2022tackling} more transparent. Various similar and complementary proposals have since followed, including nutrition labels \cite{holland2018dataset}, data statements~\cite{bender2018data}, data cards \cite{pushkarna2022data}, data briefs \cite{fabris2022tackling}, model info sheets \cite{sayash}, and causal datasheets \cite{butcher2021causal}. A common thread is the desire to standardize practices for qualitatively summarizing datasets, increasing transparency and addressing bias by exposing key details to stakeholders such as a dataset’s motivation, provenance, composition, and maintenance. As \emph{Datasheets for Datasets} \cite{gebru2021datasheets} only represents an all-purpose starting point, more specialized documentation has been proposed, for example focusing on art \cite{srinivasan2021artsheets} and health datasets~\cite{rostamzadeh2022healthsheet}.

As the creation of dataset documentation is primarily a reflective process (i.e., the data already exists), others have instead focused on the actual process of collecting data. For example, Hutiri et al. \cite{hutiri2022design} present guidelines for designing speaker verification evaluation datasets, addressing limitations in previous datasets, i.e., evaluation bias, unrepresentativeness, and unaccounted-for sources of error. LeClair and McMillan \cite{leclair2019recommendations} provide a set of recommendations alongside a new dataset, motivated by conflicting results in the task of source code summarization, which was due to a lack of community consensus on how datasets should be designed and collected. Similarly, in the library domain, despite the major role that library-linked data plays in retrieval, there is a lack of agreed upon methodological guidelines for publishing library linked data, which has now been addressed by Vilazzon et al. \cite{villazon2011methodological}.

Focusing on SLT, in addition to the guidelines of speaker verification evaluation datasets by Hutiri \cite{hutiri2022design} and the survey of Mahajan et al. \cite{mahajan2021need}, Feng et al. \cite{feng2022review} provide a review of issues related to speech-centric trustworthy machine learning, including discussing the importance of fairness and privacy for developing speech datasets. Nonetheless, no prior work focuses on the decisions speech dataset creators should make when facing speech-specific issues and appropriate documentation for such decisions, a gap that our study aims to bridge.  

\section{Speech Preliminaries}

\subsection{Speech Definitions}

A fundamental characteristic of \emph{language} is that a single meaning can be expressed in multiple different ways; some go so far as to argue that language uniformity is a pure myth \cite{evans_levinson2009}. As such, linguistic variation is detectable in all communities, even if only a single language is spoken. This variation, once noted as ``orderly hierarchy'' \cite{weinreich1968}, is what the members of the speech community use to construct both their personal and social identities, and is therefore crucial to understanding how to rightfully represent the community in an SLT system.

While it is common for speech datasets to specify which language or languages are spoken, there are more granular linguistic categorizations that can, and often should, be used to better foster speech diversity. Within a language, it is common to define a certain way of speaking as ``standard,'' which problematically erases the many other valid ways of speaking that language \cite{nakamura2019my, feng2021quantifying}.
Specifically, populations may speak with different accents, dialects, or varieties of a single language. Furthermore, individuals may speak in atypical manners, such as with speech impairments, which can manifest in different speaking styles~\cite{chlebek2020comparing,tomanek2021device,feng2021quantifying,guo2020toward,geng2022speaker,ji2022asrtest}. 
We refer to speech types along these dimensions as \textbf{``linguistic subpopulations''} and provide formal definitions for each, though we note that ongoing debate on these topics continues in many subdomains.

An \textbf{accent} is defined by Edwards \cite{edwards1997} as a unique mode of sound production affected by the speaker's specific linguistic characteristics---especially phonology and prosody. Accent is commonly thought to be a phonological variant of a language, and it is likely that certain speech features, such as the speakers' first language's phonology system, could have carried on to create accented speech \cite{lippigreen1997}. Examples of accents include non-native speech such as Chinese-accented Korean or French-accented Spanish, or regional speech such as Texan-accented English.

A \textbf{dialect} is generally considered to be more broad, and occurs when a subgroup of a language's speakers is isolated from the rest of the population \cite{edwards1997}. This would not only include phonology and prosody (as in the concept of accent), but also syntax, semantics, morphology, and pragmatics. A looser definition of dialect per Paul \cite{paul1995} is simply a version of a language that is similar to the form spoken by the majority group but different in some aspects. Our earlier example of Texan-accented English would hence classify as a dialect as well, as part of a larger Southern dialect of American English. Additional examples of regional dialects include British English or Australian English; there are also social dialects, such as African American Vernacular English (AAVE).

Language \textbf{variety} is a more neutral term that is used to refer to any type of language, encompassing all of dialect, accent, and the specific language in general \cite{trudgill2003}. Variety can include registers, styles, and idiolects, in addition to accents or dialects mentioned above. For example, Texan-accented English or African American Vernacular English would also be considered varieties of English. Additional examples of varieties that are neither dialects nor accents include honorific speech, child-directed speech, or code-switching. \textbf{Code switching} refers to a situation in which the speaker alternates between two or more linguistic varieties \cite{gardner2009code}, whether intentionally or unintentionally. Code switching can happen within a clause (intra-sentential, e.g. ``quiero ir al [es] \textit{mall next Tuesday} [en]'') or outside a clause, e.g. ``es difícil encontrar trabajo estes dias, [es] \textit{you know?} [en]''; it can also occur at a clause-level (inter-sentential, e.g. using an English phrase when speaking Spanish), or even within a single word, at a morphological level (intra-word switching, e.g. using a single English word in a sentence when speaking Spanish). 

\textbf{Speech impairments} refer to communication disorders of people with difficulties in formulating normal speech sounds necessary for communicating with others. These are often caused by disorders such as aphasia, stuttering, and lisping \cite{sheikh2021}. ASR models have been deployed for the detection of pathological or psychological issues including dysarthria \cite{millet2019learning,chen2020enhancing,shor2019personalizing}, Alzheimer's \cite{li2022alzheimer}, obstructive sleep apnoia \cite{blanco2011analyzing}, and other health concerns \cite{zhao2020hierarchical,chlebek2020comparing,lopez2017depression,clapham2012nki,lee2016automatic,kushalnagar2012readability,guo2020toward,casanueva2016improving,sunder2022building,lemmety2000review}.

A speech corpus could contain speech \textbf{uttered} by any number of linguistic subpopulations. However, this speech can present itself very differently depending on the kind of speech material used for corpus collection. The two main types of speech material are read speech and spontaneous speech. \textbf{Read speech} provides participants with a certain prompt to read aloud. When reading a text, people have a tendency to read in a more tense and formal manner, controlling for their articulation, which makes it differ from their original way of speaking \cite{mehta1988detection}. In contrast, \textbf{spontaneous speech} is when participants are allowed to speak freely, such as in a monologue or a dialogue, where the topic of the talk may or may not be predetermined. 

\subsection{Diversity of speech data}

Diversity in speech itself is multifaceted: within a dataset, it can refer to
(a) the overall distribution of languages, (b) the distribution of a single language's varieties, and (c) the distribution of speaker socioeconomic and demographic factors.

In the first case, diversity may be lacking even if a high count of languages are present in the data. Indeed, in the broader field of Natural Language Processing, there exists the concept of ``low-resource languages'' that are systematically under-represented \cite{magueresse2020low}. For example, in one of the most recent and advanced ASR models \cite{Radford2022}, some languages take up an exponentially smaller share of the underlying training data.

In the second case, diversity may still be lacking even within the realm of high-resource languages, if only the majority dialect or accent is included in the data---e.g., only including speakers of ``Standard English'' and excluding speakers of ``African American Vernacular English''~\cite{Koenecke2020}.     

This is directly related to the third case, wherein a single language can be divided into numerous diversities (usually represented as dialects or accents), often interacting with socioeconomic and demographic factors such as nationality, gender, age, education, and income level---arising in speech data not just through the human speech uttered, but also through background noises being recorded~\cite{casey2017noise, Carrier2016RoadTN}. 
Multilingual or code-switching speech is also an important component of multi-ethnic communities, and groups of people may portray distinct speech styles associated with various factors like profession. Furthermore, health status can play a role, such as for people with different impairments that take a toll on speech abilities (e.g., cerebral palsy \cite{nakamura2019my} or Alzheimer's \cite{li2022alzheimer}). 
Finally, demographic diversity in speech datasets is important to consider. While many datasets contain gender labels \cite{boito2022study,garnerin2019gender}, very few include non-binary individuals (only 3\% in the sample of datasets we reviewed, e.g., \cite{german2022spectrum,ardila2019common}). 
The same applies for speech data on different speaker age groups \cite{yu2021slt,geng2022speaker}: datasets may aggregate age into coarse groups, but the result is that certain age groups are still systematically under-represented, such as young children~\cite{yeung2018difficulties}.

Great strides have been made in each of the three cases described above. For example, there are specific datasets on spontaneous and colloquial speech for low-resource language \cite{moisio2022lahjoita}, accented speech \cite{shi2021accented}, bilingual individuals \cite{johnson2020spice}, and speech impaired individuals \cite{macwhinney2011aphasia}. That said, numerous studies have found the data lacking for all three cases: most text-to-speech services can only support a dozen languages out of the over 7,000 languages in the world \cite{tan2021survey}; and, data is still sparse for dialects \cite{dorn2019dialect,martin2021spoken} and individuals with disabilities \cite{casanueva2016improving}. Beyond the three facets of speech diversity described above, there are other dimensions to consider: there are also often limited data on domain-specific vocabulary \cite{georgila2020evaluation}, speaker geography \cite{gorisch2020using}, and, relatedly, political alignment \cite{dichristofano2022performance}.

Machine learning models trained on non-inclusive datasets further perpetuate the lack of representation, especially as in-the-wild performance is particularly dependent on the level of acoustic match with the training data \cite{karanasou2017vectors,shen2022improving}. Prior studies have found biases---based on lower ASR model performance---for speech subgroups across speech impairment \cite{markl2021context, green21_interspeech, HidalgoLopez2023}, and intersectional speech subgroups: of gender and race \cite{Koenecke2020}, gender and geography \cite{Tatman2017}, and gender and accented chatter noise \cite{walker2022biashacker}.

We suggest that practitioners be aware of the many facets of speech diversity, and be able to support their decisions to include or exclude different linguistic subpopulations in the collection process. In this way, we bring awareness to the lack of content validity \cite{Jacobs2021} in datasets that can result in representational harm \cite{Blodgett2020}.

\section{Augmented Datasheets for Speech Datasets}\label{sec:examples}

In this section, we showcase the questions included in the Augmented Datasheets for Speech Datasets. We follow the same section ordering as the original Datasheets for Datasets paper~\cite{gebru2021datasheets}. We refer to ``linguistic subpopulations'' as subpopulations speaking any of different: languages, accents, dialects, varieties, and/or impaired or atypical speech. For each question, we provide in-depth examples about how these questions can support dataset creators in ethical decision-making and enhancement of dataset transparency. 

These examples are the result of a literature review we performed on a set of 178 speech studies related to fairness and diversity, as well as 220 speech datasets.\footnote{The studies were extracted from ArXiv, The ACM Digital Library, Google Scholar, the ACL Anthology, IEEE Xplore, or were published in the following venues: INTERSPEECH, ICASSP, NeurIPS, and ICML. The speech datasets were extracted from on \url{paperswithcode.com} \cite{stojnic2022papers}, \url{huggingface.co} \cite{wolf2019huggingface}, and \url{openslr.org} \cite{openslr}. See more details on the literature review in Appendix \ref{litreview}.} For the 178 studies, we tabulated the authors' ethical considerations related to diversity, inclusion, privacy, user empowerment, crowdworker protection, data quality assessment, explainability, and the context of application of a speech technology.  For the 220 speech datasets, we investigated their diversity, inclusion, and privacy considerations, and identified limitations regarding ethical dataset development. Our literature review allowed us to map extracted information from datasets and research studies to the corresponding categories of our augmented speech datasheets, and develop representative datasheet questions for each category. 

While the ``Augmented Datasheets for Speech Datasets'' questions in this Section are each followed by additional context based on our literature review, we additionally created .tex and .docx templates for use by dataset creators and users. We include both blank templates to be filled out, as well as worked datasheet examples on common speech datasets, at \emph{\url{https://github.com/SonyResearch/project_ethics_augmented_datasheets_for_speech_datasets}}.

\subsection{Motivation}

The reasons motivating the creation of a dataset and their consequences should be well-documented. Depending on the task, or the existing gap the dataset seeks to fill, the dataset creators' motivations may influence many parameters. These may include the composition, naming, licensing terms or the data collection process.  Below, we showcase the importance of three augmented speech datasheet questions regarding motivations for dataset creation.

\subsubsection{What is the speech dataset name, and does the name accurately describe the contents of the dataset?} 
It is important to choose a descriptive and informative dataset name, ensuring it accurately describes its content. For example, it is ideal to include the speech technology a dataset is developed for (e.g., IISc-MILE Tamil ASR Corpus \cite{pilar2022subword}---where ASR refers to Automated Speech Recognition; LibriTTS corpus \cite{zen2019libritts}---where TTS refers to text-to-speech),  mention the exact demographic information it concerns (e.g., Samr{\'o}mur Children 21.09 \cite{carlosmena2021}, MAGICDATA Mandarin Chinese Conversational Speech Corpus \cite{yang2022open}, Korean Read Speech Corpus \cite{deeplyinc}, Parkinson Speech Dataset with Multiple Types of Sound Recordings Data Set \cite{sakar2013collection}), mention specific linguistic features of note (e.g., A Mandarin-English Code-Switching Corpus \cite{li2012mandarin}, Multi-dialect Arabic Speech Parallel Corpora \cite{almeman2013arabic}, Korean English Learners’ Spoken Corpus \cite{jung2021}), or provide information about the related speech domain of the recordings (e.g., Audio-Visual Speech Corpus in a Car Environment \cite{lee2004avicar}---recordings from noisy automobile environment; Deeply BibleTTS \cite{meyer2022bibletts}---recordings of Bible readings). Given that it is a standard technique to combine speech resources from different domains, languages, varieties, and styles (e.g., \cite{Radford2022}), an appropriate name selection and detailed description can assist the efficient retrieval of datasets that can lead to more inclusive models.

\subsubsection{Can the dataset be used to draw conclusions on read speech, spontaneous speech, or both?} 
A dataset focusing on speech synthesis for audiobooks would ideally contain read speech (e.g., \cite{ardila2019common,robinson1995wsjcamo,panayotov2015librispeech}), while a dataset used for in-the-wild speech recognition would contain spontaneous speech (e.g., \cite{pitt2005buckeye, ksponspeech2020}). In the case of the creation of a general purpose speech model, it would be ideal to use a dataset that contains both read and spontaneous speech (e.g., \cite{o2021spgispeech,chen2021gigaspeech}). Especially since the robustness of speech technologies such as ASR are largely dependent on their training data, it is important to guide researchers in using appropriate speech data types.

\subsubsection{Describe the process used to determine which linguistic subpopulations are the focus of the dataset.} We ask for disclosure on the process behind linguistic subpopulation choice: e.g., a dataset aiming to achieve language preservation should contain resources that focus on under-resourced languages~\cite{kuhn2020indigenous}, and the selection of content will ideally correspond to the vocabulary and culture of the corresponding community. 
In commercial speech product audits, it may be important to collect observations spanning specific parameters (e.g., accent, gender, race, user group) reflecting the diversity of the user population (e.g., \cite{fenu2020exploring,liu2022towards,meyer2020artie}).

\subsection{Composition}

We propose necessary questions that scrutinize the sufficiency and coverage of a dataset in terms of quantity (questions 3.2.1---3.2.2), quality (questions 3.2.3---3.2.7), and content of speech (questions 3.2.8---3.2.10).

\subsubsection{How many hours of speech were collected in total (of each type, if appropriate), including speech that is not in the dataset? If there was a difference between collected and included speech, why? E.g., if the speech data are from an interview and the dataset contains only the interviewee's responses, how many hours of speech were collected in interviews from both interviewer and interviewee?} 

This quantification can help ensure the robustness of the data collection process by scrutinizing choices of the dataset creators. It is important to understand why specific segments of the data are retracted, and how they handle data-related obstacles. 
For example, in MLS \cite{pratap2020mls}, the authors had to restrict the final dataset coverage given technical inabilities in retrieving book transcriptions. The VOICES dataset \cite{richey2018voices} was created by using a gender-balanced subset of another dataset, LibriSpeech \cite{panayotov2015librispeech}. Similarly, the creators of ParliamentParla \cite{kulebi2022parlamentparla} chose specific segments to optimize for vocabulary diversity. Such choices need to be carried out with caution towards robust \& inclusive dataset design and documentation. 

For datasets where interviews were conducted, the creators may need to decide whether or not to include the interviewers' speech in the raw dataset, since it will greatly affect the dataset size and properties (e.g., the interviewer's questions may be considered read speech and not spontaneous speech). Some datasets decide to exclude the interviewer speech through means such as silence detection, using the fact the interviewers are usually recorded at a much lower frequency \cite{zissman1996automatic}. Other datasets do not transcribe interviewer speech 
\cite{pitt2005buckeye}, or do not record the interviewer at all \cite{lyu2010seame}. On the other hand, some datasets choose to retain their interviewer speech~\cite{westerhout2006codas, macwhinney2011aphasia}.

\subsubsection{How many hours of speech and number of speakers \& words are in the dataset (by each type, if appropriate)?} 
We suggest describing both the number of hours included in the dataset for each combination of type (e.g., by speakers' intersectional identities) 
as well as the corresponding counts of speaker and vocabulary diversity.

Quantifying dataset size in hours can provide guidelines for model training using the data at different sizes. 
For example,  Mozilla suggests that 10,000 validated hours of speech per language is the ideal number to build an ASR system in production \cite{ardila2019common}, while for pre-trained foundational models, Zelasco et al. \cite{zelasko2020sounds} and Microsft Azure \cite{microsoft} suggest that fewer than 20 hours of target language fine-tuning 
can significantly reduce ASR error rates. Some datasets are small enough such that they can only be used for model evaluation and not training (e.g., VIVOS \cite{luong2016non}).

It is similarly important to tabulate the numbers of speakers since speaker variety can be a decisive factor for the usage of a dataset. For example, in the field of speech synthesis---more so than the speech recognition field---an important parameter is whether the dataset is collected from a single speaker \cite{park2019css10} or multiple speakers. Hence, it is crucial to pinpoint the number of speakers as well as the variability between the speakers.

Finally, it is useful to quantify the number of total and unique words being uttered. Each language has a different size of unique words, which results in different ratios of Out-Of-Vocabulary (OOV) words for the same dataset size when performing an unstratified train/test set split  \cite{creutz2007morph}. Lexicon vastness and its corresponding phonetic inventory plays a pivotal role in the performance of the ASR model, 
as OOV words often cause misrecognition of neighboring words \cite{adda2000use, woodland19951994}. Furthermore, words that do not appear as frequently as others are commonly considered outliers by the models, resulting in disparate treatment of scarce speech that many times is uttered by minority linguistic subpopulations.

\subsubsection{Are there standardized definitions of linguistic subpopulations that are used to categorize the speech data? How are these linguistic subpopulations identified in the dataset and described in the metadata?}

While it is important for datasets to be robust across different language varieties, oftentimes the  documentation is lacking in formal definitions of the linguistic subpopulations speaking in the dataset. The creators should not simply assume that a certain definition is broadly agreed upon by the linguistic community; rather, it is helpful to provide detailed definitions of such subpopulations.
This is especially true for datasets created for dialect identification or classification, or non-native speech assessment \cite{bird2019accent, boril2012arabic, ten2000asr, ahamad2020accentdb, dichristofano2022performance, speechocean762}.

The linguistic groups to be identified and described should contain recording-specific sub-groups that emerge as part of the dataset. Linguistic groups often include categories that are fluid and non-deterministic (e.g., age, gender, social background, accent, dialect). For such categories, dataset creators should make aggregation choices that speak to the diversity of the data, while respecting the privacy of the speakers' identities and histories. 
Dataset creators should explicitly describe how they identify and define such subpopulations in the dataset, while considering the implications of and absence of such categorizations.

\subsubsection{For any linguistic subpopulations identified in the dataset, please provide a description of their respective distributions within the dataset.}

It is important to understand the representational makeup of speakers and utterances in a speech dataset, in particular among linguistic subpopulations, and among socioeconomic and demographic groups. Different levels of representation in speech data used to train models can have different downstream effects on model performance among diverse test sets. Augmenting data to make the original training dataset more balanced in terms of gender has been found to yield improvement in model performance \cite{choubey2021improving}, and there have been attempts to create datasets specifically tackling gender balance in the training data for fields with known imbalances \cite{webster2018mind}.
In our review, we found that fewer than 10\% of the datasets were balanced on demographic properties of gender and age. That said, we note that ensuring demographic parity in \emph{all} datasets would disproportionately penalize researchers for generating important training data focused on underrepresented minority language types that are time- and labor-intensive to collect.

\subsubsection{How much of the speech data have corresponding transcriptions in the dataset?}

It is useful to first understand whether speech transcriptions are necessary for the SLT tasks at hand, and if so, to what extent they are provided.
For example, a model to simply detect background noise would not require transcriptions---it would only need labeling of spoken noise.
Meanwhile, training an ASR model necessitates having the ground-truth transcription provided with the speech dataset (due to the underlying language model using textual data), and will yield better performance with high quality transcriptions \cite{bu2017aishell}. 
However, the cost of generating a high volume of transcriptions for large speech datasets is extremely demanding, so current ML methods such as semi-supervised learning or weakly-supervised learning have been developed to allow for training on only a small initial transcribed dataset, and then self-learning to make hypotheses on the remaining larger portion of the unlabeled dataset \cite{ling2020deep, Radford2022, li2022combining}. Furthermore, dataset creators may use existing ASR models to generate transcriptions of speech recordings---significantly less expensive than hiring human transcribers---which can inject pre-existing biases of the ASR model into a new speech dataset (\cite{zuluaga2022atco2, lakomkin2019kt}). Another technology used is data augmentation, where an in-domain, fully-transcribed dataset is adapted---mainly through acoustic modification---to be used as training data to predict the ground truth transcriptions of a new dataset \cite{kumar2022creating, kadyan2022prosody}. It is important to enumerate transcriptions and state their limitations in a dataset so practitioners can consider whether these enhancement methodologies can suffice for their SLT.

\subsubsection{Does the dataset contain non-speech mediums (e.g. images or video)?}
Oftentimes, speech data are extracted from videos crawled from public domains \cite{takamichi2021jtubespeech, bougrine2017toward, gretter2014euronews}.
Furthermore, there are specific use cases requiring multimedia datasets, such as discourse analysis of speech with a high level of difficulty in voice intelligibility \cite{macwhinney2011aphasia}. Relatedly, combining audio and visual resources has been found to improve ASR performance in low speech-to-noise settings \cite{lee2004avicar}. 
Finally, there are use cases where additional data formats can benefit ML prediction tasks. 
For example, the Turkish Audio-Visual Bipolar Disorder Corpus \cite{cciftcci2018turkish} uses audio-visual data because the creators' goal is to recognize affective states of patients for bipolar disease classification by combining both audio and appearance features. As such, it is useful to consider whether additional media formats should be released with the speech data.

\subsubsection{Do speakers code switch or speak multiple languages, and if so, how is this identified in the data?}
One of the main sources of errors in ASR is words being uttered in different languages \cite{wirth2022asr}. Hence, it is important to document when code switching happens, ranging from single words, to sentences or full segments of speech recordings \cite{mubarak2021qasr}. Especially since data-subjects may not be aware  that they are code-switching, dataset curators should carefully review speech and include  appropriate label tags to indicate a change of linguistic feature.
Code-switching may be more difficult to transcribe, since languages used to code-switch may use different alphabetical systems, and each language could require separate annotation systems (e.g., TALCS---using Mandarin and English \cite{Li2022TALCSAO}). Some code-switching corpora include a separate layer of transcription to mark where the code-switching happens, either inter-sentential or intra-sentential (e.g., TuGeBic \cite{cetinoglu2017code}).

\subsubsection{Does the speech dataset focus on a specific topic or set of topics?}
It is important to carefully choose and document the \emph{content} of speech included in the dataset. 
For example, speech focusing on finance \cite{o2021spgispeech} or politics \cite{kirkedal2020ft} may not be suitable for training general-purpose synthesis or recognition systems. In addition, specific applications for speech systems such as air traffic control \cite{yang2019atcspeech} or the medical domain \cite{koreanmedical}
contain disproportionate amounts of domain-specific vocabulary and jargon, which may be out of scope in pre-trained general purpose models~\cite{georgila2020evaluation}. 
Therefore, it is crucial for dataset creators to report the content scope of the speech recordings. 

\subsubsection{Does the dataset include sensitive content that can induce different emotions (e.g., anger, sadness) that can cause the speakers to produce unusual pitch or tone deviating from plain speech?} 
Datasets that contain only emotion-neutral speech \cite{shi2020aishell} might be less useful for broadly deployed ASR systems, or for speech synthesis models for music or other creative tasks. In addition, emotional language diversifies prosodic \cite{Frick1985} and phonetic elements of language \cite{aguiar2014voce}, resulting in lower performance in models \cite{ji2022asrtest}. Thus, it is helpful to include---and perhaps label---emotional speech in datasets.

\subsubsection{Does the dataset contain content that complies to the users' needs, or does it result in symbolic violence (the imposition of religious values, political values, cultural values, etc.)?}

The problematic composition of a dataset can lead to the creation or perpetuation of power asymmetries in the society, a term that sociologist Bourdieu terms as \textit{symbolic violence}---defined as a type of non-physical violence manifested in the power differential between social groups \cite{bourdieu1990reproduction}.  Unless the dataset is specifically aiming to collect a certain portrayal of reactions to power dynamics, it is the creator's responsibility to attempt to discard content deemed as promoting symbolic violence \cite{mackenzie2020cc}. 
For example, the usage of the Bible or highly value-laden text (e.g., novels with colonialist themes \cite{bird2020decolonising}) for creating speech recordings should be avoided among some communities, due to the differences in cultural \& religious values and histories---being forced to reach such texts could exacerbate the powerlessness felt by those communities \cite{teodorescu2022cree, meyer2022bibletts}. Furthermore, the mere presence of certain kinds of speech in a released dataset could be enough to perpetuate symbolic violence towards dataset users---regardless of whether or not the dataset creators condone such speech. 

It is therefore important to document whether such limitations and issues exist in a dataset and also strive to focus on dataset content that complies with users' needs and respects data subjects. The needs of data subjects can be taken into consideration by focusing on recordings of vocabularies, topics, and speech segments that best represent subjects' every day lives, values, social and personal behaviors.

\subsection{Collection Process}

Speech data collection encompasses processes that could involve the scraping of digital or analog content, and/or the cooperation with data-subjects who will provide their voice/recordings.

\subsubsection{What mechanisms or procedures were used to collect the speech data, e.g.: is the data a new recording of read speech, or an interview? Or is it downloaded speech data from public speeches, lectures, YouTube videos or movies, etc.?}

Dataset creators should strategically and carefully decide on the mechanisms they use to collect speech. For example, when scraping online content \cite{hernandez2018ted,neto2011media,irie2018radmm,hessel2020beyond,kearns2014librivox}, they should ensure that they have permissions to acquire content, and disclose the appropriate permissions \cite{liu2010very,snyder2015musan,feng2022review}. Furthermore, they should describe whether the speech is a direct recording, acquired from a pre-existing speech source, or if they interviewed data-subjects specifically to create this dataset. Note that additional questions on interviewee recruitment and consent are covered in the original ``Datasheets for Datasets'' paper \cite{gebru2021datasheets}.

\subsubsection{Were all the data collected using the same technical methodology or setting, including the recording environment (e.g., lab, microphone) and recording information (e.g., sampling rate, number of channels)?}

The use of different speech collection media can have a significant impact on the quality of collected speech.
Relevant properties include the distance of individuals from the microphone, the type of microphone \cite{irfan2020challenges}, single or multi-channel data collection \cite{liu2016sheffield,morales2008stc,yu2021slt}, and recording frequency. For example, speech recordings of higher frequency than 20 khz \cite{adiga2021automatic,stan2011romanian} might be preferred over lower ones \cite{snyder2015musan,o2021spgispeech} in speech synthesis, since they are of higher quality and improve user experience. Similarly, data collection using only specific types of microphones \cite{stan2011romanian} could reduce hardware-induced noise in recordings that used random or not model-specific recording media \cite{du2018aishell,karpov2021golos,qader2019kurdish,michailovsky2014documenting,harveenchadha,li2021oriental,Radeck-Arneth2015,janin2003icsi}.
Additionally, in cases of collection crowdsourcing, dataset creators should reflect on what technical requirements and processes they set for collecting speech, since individuals might not have uniform internet access \cite{abraham2020crowdsourcing}, or be familiar with complicated user interfaces \cite{kuhn2020indigenous,lata2010development}.

\subsubsection{Is there presence of background noise?}

On spontaneous speech datasets, variable background environments can result in differential performance of speech models \cite{gorisch2020using,chen2021gigaspeech}. 
ASR model performance declines with a smaller the signal-to-noise ratio \cite{martin2021spoken,robinson2000speech}. Generally, depending use case (e.g., speech synthesis), or dataset motivation (e.g., language preservation), researchers might prefer recordings in quiet environments \cite{stan2011romanian,shinodatokyo,halabi2016modern}, or enhancing recordings with artificial noise \cite{cosentino2020librimix,veaux2017cstr}.  However, using datasets that are highly varied in noise levels---from diverse environmental conditions---can result in more robust models~\cite{ardila2019common,barker2018fifth,gorisch2020using,georgila2020evaluation,black2011automatically,ji2022asrtest,kawakami2020learning, microsoft_research_2022}.

Noise-related considerations are not merely for the sake of building robust ASR models (on its own, a long-standing task in the field \cite{ris2001assessing, maas2012recurrent, feng2021asr}); noise levels also have direct socioeconomic implications, as they are estimated to be higher for population groups with higher proportions of lower-socioeconomic status residents, with factors such as race, poverty, unemployment, and education level having impact \cite{casey2017noise, Carrier2016RoadTN}. Casey et al. \cite{casey2017noise} even considers linguistic isolation as one of the main factors correlating with higher levels of noise, where ``linguistic isolation'' is defined as a household where no one over the age of 14 speaks English ``very well.'' In order to fully consider the diversity of socioeconomic, ethnic, or racial make-up in speech data, it is critical to ensure that data pre-processing steps removing background noise do so in a principled way that does not erase representation across socioeconomic levels.

\subsubsection{For interviewer/interviewee speech data: during the interview process, did interviewers consistently ask questions that are ``fair and neutral''?}

Qualitative data collection---and in particular, via interviews---is a common method to collect speech data, especially for spontaneous speech.  
In the survey bias literature \cite{williams1968interviewer}, value neutrality is defined as being objective to personal values when conducting sociological studies and staying clear of engaging personal values and opinions \cite{weber1949objectivity}. A critical component of conducting interviews is maintaining this neutrality \cite{weiss1995interview, gerson2020interviewing}. Interviews used for such data collection are usually open-ended, meaning they cannot be answered with a simple ``yes'' or ``no'' response, and instead prompt a longer conversation, making the interview become semi-structured.
Even though the semi-structured nature of the interview makes it difficult to anticipate the response and what question to move on to, it is important for the interviewers to remain neutral at all times and withhold judgment \cite{hammer1993}.

The use of ``fair and neutral'' questions allows for open, honest, and respectful communication without judgment between the interviewer and interviewee, who may have different viewpoints and lived experiences.
Fair questions consider the context, circumstance, and perspective of the respondent. This necessitates careful consideration with regard to the impact a question can have on a respondent. Fair questions should be both respectful and considerate, reflecting that each person has their own set of opinions and experiences. In addition, the interview should make interviewees feel comfortable, both in terms of interview content and setting. For example, Pitt et al. \cite{pitt2005buckeye} balanced the interviewers and interviewees gender combinations, to control for social dynamics developed in inter- and intragender discussions. In a similar fashion, Kendall and Farrington \cite{kendall2018corpus}---for data collection of African American English speech among interviewees---also ensured that the interviewers were speakers of African American English to minimize code-switching between the two parties.

Neutral questions are neither biased or leading, thus resulting in more objective and unbiased responses. Such questions are free from assumptions and suggestive language that might conceivably prejudice a respondent on the basis of the interviewer’s personal position. Neutrality can be supported by using open-ended questions or diverse topics of discussion \cite{datatang}, in order to freely drift language usage according to their habitual frequency \cite{lee2019talking} and capture phenomena such as diction \cite{kirkpatrick2020natural}.

We include two examples of fair and neutral questions. Firstly, from Holstein et al. \cite{holstein2019improving} to machine learning practitioners:  ``Can you recall times you or your team have discovered fairness issues in your products?'' The question is open-ended, respectful of the practitioners’ product knowledge, expertise, and experience, and permits response autonomy (respondents are able to exert control over what they share and how). Secondly, from Kapoor et al. \cite{kapoor2022weaving} to labor organizers in the US technology industry: ``How do you increase membership in your collective? Are you involved in reaching out to other employees who are not yet a part of [collective’s name]?'' The question is open-ended, providing respondents an opportunity to share their strategies without any assumptions or biases with respect to how this should be done. Moreover, the question is respectful and non-judgmental—it does not assume that the respondent is, or should be, involved in outreach activities.

\subsubsection{Have data subjects consented to the disclosure of the metadata in the dataset? Also, does the metadata include sensitive personal information such as disability status?}

Basic metadata release is critical for dataset usage, but depending on the dataset's purpose, additional metadata may contain information related to sensitive or biometric properties of the person---it is important to ensure the data subject agrees to both the validity and release of such metadata. For example, metrics could include  health and speech impairment status that could negatively impact the individual if such information were made public \cite{yoon2019development,aguiar2014voce,sakar2013collection}. A common machine learning task is one of score prediction---e.g., on the Mini-Mental State Examination (MMSE)---using impaired speech corpora. Datasets collected with such a task in mind would include not just speech data, but also metadata on the participants' MMSE scores \cite{LuzHaiderEtAl20ADReSS}. Additional metadata can include detailed information about the etiology, the onset/duration time of the impairment, additional disorders, and even surgical information \cite{westerhout2006codas, lee2022building}---all of which have the potential to do serious harm to a data subject if stored improperly or released without consent. Finally, it is important to ensure that an individual's metadata is agreed upon---for example, it is helpful to have data subjects self-report demographic information (e.g., \cite{ardila2019common, turrisi2021easycall, gupta2022adima}), since it may be harmful for individuals to see inaccuracies about their identities; such inaccuracies can also affect the utility of the dataset and ability of developers to conduct fairness evaluations.

\subsection{Preprocessing/Cleaning/Labeling}

Speech datasets usually go through significant processing and cleaning, since audio segments need to be aligned and standardized, and transcriptions also often need to be annotated.

\subsubsection{When generating the dataset, was any background noise deleted or adjusted to make all recording qualities similar?}

Processing dataset noise is common for speech data. When different recording media is used, creators often convert speech to uniform formats \cite{veaux2017cstr,snyder2015musan,mohamad2019shemo}. Furthermore, there are datasets that explicitly add artificial noise for speech or perform related augmentations, especially for tasks such as speech separation or quality assessment \cite{wichern2019wham,cosentino2020librimix,reddy2020interspeech}.  Such transformations should be reported, since they influence the acoustic and time-series properties of the dataset. 

\subsubsection{Did the data collectors hire human annotators to transcribe the data? If so, how trained were the annotators in speech transcription for this context? How familiar were they with the corpus material, the vocabulary used, and the linguistic characteristics of different dialects and accents?}

The annotation of speech either takes place by automated means (i.e., an ASR model) to be later reviewed by human annotators, or is directly annotated by human annotators. In both cases, it is crucial to consider sociolinguistic factors that influence the quality of the annotations. Wirth et al. \cite{wirth2022asr} showed that more than 15\% of ASR errors can be traced back to flawed ground truth transcripts. Many times, these errors are a result of annotators not being fluent speakers of a specific dialect or accent \cite{bhogale2022effectiveness,cousse2006regional}, with false annotations being systematically found for dialects of underrepresented groups such as African American English \cite{Jones2019}. To mitigate such phenomena, creators should recruit annotators that have extensive knowledge of dialects or accents appearing in the speech corpus, and are familiar with the vocabularies included. Furthermore, many languages and dialects contain features that are exclusive (e.g. certain vowels \cite{park2019jejueo}), are highly agglutinative \cite{teodorescu2022cree}, or even do not have a written form. In such cases, it is important to define guidelines on how to transcribe speech \cite{moisio2022lahjoita}, drawing from existing consortium guidelines \cite{pitt2005buckeye,liu2010very,boito2022trac}  or linguistically motivated rules \cite{mubarak2021qasr}. Dataset creators should explain how annotators were recruited and trained, including training in and use of transcription tools like Praat \cite{praat}.

\subsubsection{If multiple transcription methods were used, how consistent were the annotators? How were transcripts validated?}

Transcriptions can be unreliable and subjective, so it is important to document the methodology for generating them. Prior research has shown that annotators do not always agree on transcriptions \cite{black2011automatically}, especially in tasks that are highly subjective \cite{engelmann2022people}, such as naturalness, coloration, discontinuity, loudness, and emotional expression \cite{reddy2020interspeech,hu2007subjective,feng2022review,sharma2021survey}.
Even in simpler transcription cases annotators might disagree, since language is never perfectly neutral, nor does it convey a single meaning \cite{wagner2019speech}. Therefore, it is important to implement appropriate inter-annotator scoring systems \cite{pitt2005buckeye, brants2000inter, valenta2014inter} that vote on the final annotations and labels of the dataset. Furthermore, appropriate guidelines should be provided that clarify how annotators should label in ambiguous cases \cite{mohamad2019shemo}, together with filtering \cite{das2016automatic}, alignment \cite{abraham2020crowdsourcing}, and qualitative error analysis processes \cite{markl2021context} that validate the final annotations. Depending on language and use cases, such processes can be expanded by practices that increase diversity, such as usage of versatile transcriptions for language without standard orthography \cite{hussein2022arabic,ali2017speech} and scripting techniques that adapt to annotators' accents \cite{teodorescu2022cree}. 

\subsubsection{If the speech data include transcriptions, what software was used in the generation of the transcriptions (including, e.g., software used by human transcribers)? Are timestamps included in transcriptions? Are the alignments provided with the transcripts?}

When a speech dataset includes human-generated transcriptions, creators should carefully choose which software to use and why, as it may interfere with  transcription quality. The most commonly used tools for linguistic annotations are ELAN \cite{elan}, CLAN \cite{clan} and Praat \cite{praat} (Appendix \ref{software}), which are software programs created for linguistic annotation purposes. They have distinct methods of saving transcription files, and provide multiple methods to auto-segment the audio files according to the creators' needs. ELAN and CLAN provides video annotation options, making it possible to annotate multimodal datasets \cite{eijk2022cabb, pengcuo2021research} while Praat is used more often for phonetic detailing \cite{gronnum2009danish, salesky2020corpus}.  When the transcription is done manually through a basic word processing tool and needs to be aligned with the speech data, there are forced alignment tools available (e.g., \cite{mcauliffe2017montreal, gorman2011prosodylab}), given the appropriate ground truth transcription. Creators can also build their own aligner (e.g., \cite{cavar2016generating}); however, since each aligner can yield different outputs and may cause different errors, this leads to dissimilarity in performance levels (\cite{mahr2021performance, gonzalez2020comparing}). Timestamps are important, since it is the information used to extract utterances, and on what level the timestamps are labeled also play a part in the alignment of the transcription to the audio data. Inaccurate timestamps will truncate the speech or mix fragments with neighboring ones~\cite{wang2021voxpopuli}.

When transcriptions are ASR-generated as a first pass, creators often use commercial or open source speech-to-text services that are publicly available (e.g., \cite{ahmed2020preparation}). However, humans still need to manually check for errors in the final stage.

\subsubsection{Were transcription conventions (such as tagging sche-me, treatment of hate speech or swear words, etc.) disclosed along with the corpus?}
 
Because there are multiple transcription conventions available to dataset creators, it is thus difficult for users to know how a transcription is generated without additional information. Even datasets managed by the same organization can have different conventions.\footnote{For example, even though they were both created by the Korean government, KSponSpeech \cite{ksponspeech2020} and AI Hub \cite{aihub} both provide the orthographic and phonetic transcription of a single word if the speaker enunciates in a non-standardized way, but the ordering is different, requiring the users to pre-process the transcription in two different ways.} Even if the creators decide to use a well-known convention such as the Text Encoding Initiative (TEI) \cite{tei} or the CHAT standard \cite{macwhinney2011aphasia}, or decide to follow a transcription methodology used by a widely used corpus like Librispeech \cite{panayotov2015librispeech} or Common Voice \cite{ardila2019common}, it is still necessary to provide a detailed transcription convention along with the actual dataset. Some necessary details to include are: where to put utterance boundaries; how to treat incomplete or unintelligible speech; how to standardize foreign words or loanwords; and whether and to what extent non-lexical backchannels (e.g., filler words) should be included in the regular transcript \cite{sadowsky2022sociolinguistic}. As any form of hate speech or swearing may occur in spontaneous speech collection, the onus is also on the creators to decide whether they will redact those terms or leave them in the transcription. Commercial speech-to-text products provide the option of profanity filters as well (e.g., \cite{googleprofanity}). An additional set of conventions to consider is tagging, which is done for specific types of information the dataset intends to deliver. This includes grammatical tagging of part-of-speech \cite{santos2014corpus,johannessen2007advanced}, emotion tagging \cite{devillers2002annotation, sini2018synpaflex,banga2019indian}, and audio event or type tagging \cite{gemmeke2017audio, thompson2010building, liao2020formosa}.

\subsubsection{Is additional coding performed, separate to transcriptions and tagging?}

In case of special coding done in addition to the textual transcription or tagging, there is no conventional systematization among dataset creators. Some datasets include information such as background noise in their transcription---even specifying which types of noise they are \cite{macwhinney2011aphasia,ksponspeech2020, barker2018fifth}. Others simply include the basic orthographic and/or phonetic transcription. There may also be additional linguistic information coded, depending on the specialty the dataset is focusing on, such as discoursal, stylistic, or pragmatic information \cite{garside1997corpus}. For example, the Hong Kong Corpus of Spoken English \cite{cheng2005creation} has prosodic features such as tone boundaries, prominence and intonation types, represented in the data. While such annotation conventions are fairly common in linguistic research \cite{cauldwell2002, brazil1997}, this is not the case in the broader speech dataset community, so detailed explanations should be provided. 

\subsection{Uses/Distribution/Maintenance}
Since the purpose of speech datasets depends on whether they are for public, commercial, or private use, we pose questions can inform the possibilites and limits of their use, distribution, and maintenance.  

\subsubsection{How are redactions performed on the dataset? Are personally identifiable information or sensitive information removed from only transcripts, audio censored from the speech data, or both?}

A critical component of dataset creation is the redaction\footnote{The United States National Institute of Standards and Technology defines several terminologies in the field of redaction: personal or identifying information is used to indicate information that is either from individuals or used to identify a certain individual. De-identification or anonymization are terms used to refer to the process of removing the association between the representation in the dataset and the represented individual, and redaction indicates the process of removing any personal or sensitive information from the dataset \cite{garfinkel2015identification}. Though terminologically distinct in specific ways, the three terms are used interchangeably in the field of speech datasets.} of information that might violate the privacy of data-subjects or can be used for harm, especially against vulnerable populations in the data \cite{ringel2019ai}. While the release of a speech dataset allows for the broader public to gain access to data, this may create tension between the protection of privacy and the goals of transparent data use. Data redaction or de-identification can ameliorate this tension \cite{garfinkel2015identification}, and this redaction of personal information is often exemplified in the medical domain \cite{cohn2019audio}. Since anonymization and filtering techniques have limited success \cite{srivastava2020evaluating} and/or significantly lower the quality of speech \cite{wu2021understanding,lopez2017depression}, the most efficient method for both protecting individuals and maintaining quality of the data is the redaction of speech and the corresponding captions that could leak personal information to the public \cite{gupta2022adima}. In the best case, both the transcript and audio would be redacted \cite{kendall2018corpus}; the next best effort is often observed where only the transcript is redacted but not the audio \cite{macwhinney2011aphasia, aihub, ksponspeech2020}. Furthermore, privacy issues should also be addressed via informed consent. If data subjects know that what they say might become publicly available information, that can mitigate the risks that they say something they would not like to be released publicly.

\subsubsection{Is there any part of this dataset that is privately held but can be requested for research purposes?}

Many times, a specific dataset is a subset of a larger one, but the extended resource might only be available as a commercial product \cite{surfingtech}. Similarly, parts of a dataset might not be provided publicly due to privacy constraints, limited access based on profession or citizenship, or other reasons \cite{kolobov2021mediaspeech}. If it is possible to use the extended version of a dataset, it is ideal to disclose this and elucidate the corresponding access requirements. 

\subsubsection{Is there a sample dataset distributed? If so, how well does the sample represent the actual dataset? Do they include all forms of speech included in the dataset? How big is the sample?}

Dataset owners often provide a \textit{light} version of the corpus, which can be used when there are data size limitations \cite{chen2021gigaspeech}. In this case, the sampled version should ideally contain the same properties of the original sample, both qualitatively (using similar metadata and and annotation conventions) and quantitatively (containing observations of different linguistic groups at similar rates). Differences between the sampled and full dataset should be disclosed to users.

\subsubsection{Aside from this datasheet, is other documentation available about the data collection process (e.g., agreements signed with data subjects and research methodology)?
}

More elaborate repositories are often necessary \cite{moisio2022lahjoita} to contain the exact agreements signed with data-subjects and annotators, read-me files that describe metadata, audio, technical infrastructure, and documentation of the overall collection process. Such documentation can ensure the reproducibility of datasets, the identification of errors, and the reassurance of best practices in ethical dataset development.

\section{Discussion \& Limitations}\label{sec:discussion}

Our augmented datasheets provide a blueprint of considerations that should be carefully reviewed and addressed by practitioners. The benefits to dataset \emph{creators} include ensuring standardization of dataset documentation, enhancing transparency of dataset contents, clarifying the underlying motivations and process of data collection, and encouraging explicit consideration of underrepresented linguistic subpopulations and socioeconomic/demographic groups. The benefits to dataset \emph{users} include more comprehensive understanding of dataset utility, and easier decision-making on data selection for more robust and inclusive SLT---especially for data on underrepresented groups. Regardless of whether the datasheet questions are being answered by users or creators, engaging with each question can provide valuable reflexive knowledge on ethical machine learning development. We suggest that dataset creators release a completed augmented datasheet alongside their speech dataset to inform the broader research community about the dataset's possibilities and limitations; this release also sets an example for more ethical, inclusive, and transparent machine learning practices.

We encourage SLT practitioners to view datasheets as a collaborative process, engaging dataset users, data subjects, and affected communities. 
A plausible path would be to perform user-centered research before creating a dataset \cite{maharjan2022experiences} in order to understand privacy concerns of data-subjects, implement safe-guards about the metadata released, and develop corresponding agreement forms \cite{ratner2018fluency}. Simultaneously, dataset creators should develop protocols and data retention strategies through IRB or institutional ethics committees \cite{turrisi2021easycall}. Especially in cases of datasets that focus on low-resource and vulnerable communities, dataset creators should ideally consult and cooperate with local communities that can inform the ethical data collection \cite{butryna2020google,abraham2020crowdsourcing,kuhn2020indigenous,siminyu2022corpus}, by taking into consideration the geopolitical origin and properties of dialects, accents, and languages. Furthermore, it is valuable to perform in-the-wild experimentation including focus groups on developed SLT applications, which can uncover issues and harms affecting user populations~\cite{clark2019state}.

Our augmented datasheets mark an improvement over the current state of speech dataset documentation. During our large-scale literature and dataset review, we found numerous ethical questions remained unaddressed, and many dataset properties undisclosed, leading to confusion about dataset utility (Appendix \ref{datareview}). That said, there are still several limitations of our work. These limitations primarily stem from two factors: the generation of the datasheet questions, and the lack of ability to dictate how the datasheets will be used on speech datasets. Regarding the first limitation, our positionalities (predominantly as researchers) limit our perspectives. To address this concern, we performed an extensive literature review to uncover pain points emerging during practitioners' speech dataset development processes. Regarding the second limitation, we encourage datasheet use for practitioners---not just individually, but also in an updating feedback-loop between dataset creators and users. Furthermore, our datasheets can only be used as a tool, and cannot make unilateral ethical choices on behalf of practitioners in lieu of deeper conversations.

For example, in domains such as literacy assessment \cite{yoon2019development,black2011automatically,nicolao2018improved} it is expected that ASR models are \emph{only} robust to a ``standard'' conception of language, rather than encompassing more diverse, atypical speech.
As another example, models of speech recognition or synthesis in affective computing might support users to understand or express their feelings \cite{mallol2020investigation,lopez2017depression,banga2019indian,wagner2019speech,gupta2022adima}. But, dataset creators may be concerned about this work mediating serious biases when humans assess or interpret the emotional conditions of others during data labeling, potentially leading to physiognomic and pseudoscientific inferences \cite{stark2022physiognomic,engelmann2022people}. The ethical choices that practitioners must make in these use cases are ones that augmented datasheets can help with in information gathering and conversation starting, but not with final decision-making.

We also believe additional research is needed in the space, especially as emerging techniques are introduced. For example, synthetically generated data using generative AI \cite{borsos2022audiolm,agostinelli2023musiclm} could serve as a solution for expanding low-resource language datasets \cite{billa2021leveraging,li2018training}; however, this raises further ethical questions about diversity, authenticity, and robustness of included speech data. Our proposed augmented datasheets can serve as a tool to evaluate, document, and discuss such advancements applied to SLT.

\section{Research Ethics \& Social Impact}

When conceptualizing the augmented datasheets template, we focused on issues of diversity and inclusion of linguistic subpopulations, and ethical treatment of data-subjects and annotators. We did not obtain or use any private or sensitive information, nor did we recruit users or perform experiments. Our augmented datasheets framework has broader impact by increasing transparency of speech datasets and fostering ethical decision-making by dataset users and creators, who will be prompted with questions such as: the representativeness of linguistic subpopulations and socioeconomic groups in their data; the ethical treatment of speaker subjects (such as avoiding ``symbolic violence'' or release of private data); and the proper designation of linguistic features (such as referring to a speaker's dialect with a term the speaker identifies with). Practitioners can use our augmented datasheets to proactively consider social features that can mitigate biased SLT outcomes.



\bibliographystyle{acm}
\bibliography{main}

\appendix
\onecolumn

\section{Augmented Datasheets for Speech Datasets} \label{datasheets}

\subsection{Motivation}
\begin{itemize} 
\item What is the speech dataset name, and does the name accurately describe the contents of the dataset?
\item Can the dataset be used to draw conclusions on read speech, spontaneous speech, or both?
\item Describe the process used to determine which linguistic subpopulations are the focus of the dataset.
\end{itemize}

\subsection{Composition} 
\begin{itemize}
\item How many hours of speech were collected in total (of each type, if appropriate), including speech that is not in the dataset? If there was a difference between collected and included, why? E.g., if the speech data are from an interview and the dataset contains only the interviewee's responses, how many hours of speech were collected in interviews from both interviewer and interviewee?
\item How many hours of speech, number of speakers \& words  are in the dataset (by each type, if appropriate)? 
\item Are there standardized definitions of linguistic subpopulations that are used to categorize the speech data? How are these linguistic subpopulations identified in the dataset and described in the metadata?
\item For any linguistic subpopulations identified in the dataset, please provide a description of their respective distributions within the dataset. 
\item How much of the speech data have corresponding transcriptions in the dataset?
\item Does the dataset contain non-speech mediums (e.g. images or video)?
\item Do speakers code switch or speak multiple languages, and if so, how is this identified in the data?
\item Does the speech dataset focus on a specific topic or set of topics?
\item Does the dataset include sensitive content that can induce different emotions (e.g., anger, sadness) that can cause the speakers to produce unusual pitch or tone deviating from plain speech?
\item Does the dataset contain content that complies to the users' needs, or does it result in symbolic violence (the imposition of religious values, political values, cultural values, etc.)?

\end{itemize}

\subsection{Collection Process} 
\begin{itemize}
    \item What mechanisms or procedures were used to collect the speech data, e.g.: is the data a new recording of read speech or an interview? Or is it downloaded speech data from public speeches, lectures, YouTube videos or movies, etc.?
    \item Were all the data collected using the same technical methodology or setting, including the recording environment (e.g., lab, microphone) and recording information (e.g., sampling rate, number of channels)?
    \item Is there presence of background noise?  
    \item For interviewer/interviewee speech data: 
    during the interview process, did interviewers consistently ask questions that are ``fair and neutral''?
     \item Have data subjects consented to the disclosure of the metadata in the dataset? Also, does the metadata include sensitive personal information such as disability status?
\end{itemize}

\subsection{Preprocessing/cleaning/labeling}
\begin{itemize}
    \item When generating the dataset, was any background noise deleted or adjusted to make all recording qualities similar? 
    \item Did the data collectors hire human annotators to transcribe the data? If so, how trained were the annotators in speech transcription for this context? How familiar were they with the corpus material, the vocabulary used, and the linguistic characteristics of different dialects and accents?
    \item If multiple transcription methods were used, how consistent were the annotators? How were transcripts validated?
    \item If the speech data include transcriptions, what software was used to generate the transcriptions (including, e.g., software used by human transcribers)? Are timestamps included in transcriptions? Are the alignments provided with the transcripts?
    \item Were transcription conventions (such as tagging scheme, treatment of hate speech or swear words, etc.) disclosed along with the corpus? 
    \item Is additional coding performed, separate to transcriptions and tagging?
\end{itemize}

\subsection{Uses / Distribution / Maintenance}
\begin{itemize}
    \item How are redactions performed on the dataset? Are personally identifiable information or sensitive information removed from only transcripts, audio censored from the speech data, or both?
    \item Is there any part of this dataset that is privately held but can be requested for research purposes?
    \item Is there a sample dataset distributed? If so, how well does the sample represent the actual dataset? Do they include all forms of speech included in the dataset? How big is the sample?
    \item Aside from this datasheet, is other documentation available about the data collection process (e.g., agreements signed with data subjects and research methodology)?

\end{itemize}
\section{Literature Review} \label{litreview}

\begin{figure*}[htpb]
    \centering
    \includegraphics[width=1\textwidth]{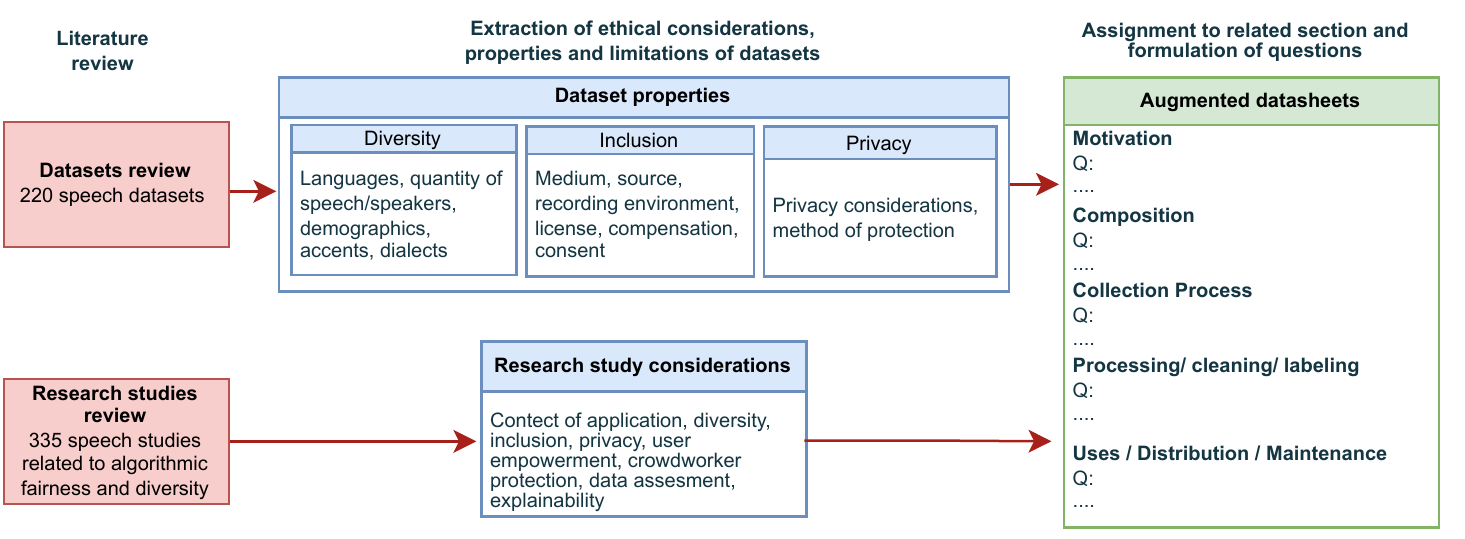}
    \caption{Overview of literature review and the creation of questions in the augmented datasheets. We reviewed 220 speech datasets and identified properties that ethically collected datasets should possess. We also identified limitations in terms of diversity, inclusion, and privacy. Furthermore, we reviewed 335 studies related to algorithmic fairness and diversity, focusing on 178 manuscripts in detail, and extracted data-centric ethical considerations. We mapped this information to the corresponding sections of the Augmented Datasheets, and formulated speech-specific questions.}
    \label{fig:litreview}
\end{figure*}
\subsection{Datasets}

We collected 432 datasets from paperswithcode.com \cite{stojnic2022papers}, huggingface.co \cite{wolf2019huggingface}, and openslr.org \cite{openslr}, all of which contain repositories of speech corpora. For each platform we extracted any dataset containing the query \textit{speech} in them. Then, we manually examined the corpus, removed duplicates, and dropped datasets that were falsely categorized. This resulted in 220 datasets, which we reviewed for best practices and issues in speech dataset development and can be found \href{https://github.com/SonyResearch/project_ethics_augmented_datasheets_for_speech_datasets}{here}. To get a more systematized overview of the datasets in terms of ethical considerations, we randomly sampled 100 of the datasets, and created a set of descriptive features we searched for, which can be aggregated in following categories:
\begin{itemize}
    \item Diversity - what different properties does a dataset posses?
    \item Inclusion - how is the dataset created and under which conditions?
    \item Privacy - how does the dataset protect the privacy of data subjects?
    \item Other.
\end{itemize}
The underlying features include information about the amount of hours in the speech corpus, the language, accent, and dialect diversity in them, technical and procedural aspects in data collection, and distributional and privacy considerations. These categories can be found in Table~\ref{datasetsoverview}.

\begin{table}
\caption{Overview of the dataset properties that we examined in a subset of 100 speech datasets.}

\begin{tabular}{ccl}
\textbf{Category}           & \textbf{Field}                      & \textbf{Explanation}                                                                  \\ \hline
 Diversity & N. languages                       & How many different languages does the dataset include?                                \\[0.1cm]
                            & N. hours                           & How many hours of speech does the dataset include?                                    \\[0.1cm]
                            & N. tokens                          & How many total and unique tokens (words) does the dataset include?                            \\[0.1cm]
                            & N. speakers                          & How many individuals were recorded in the dataset?                            \\[0.1cm]
                            & Resource                           & Are these languages low or high resource?                                            \\[0.1cm]
                            & Languages                           & What languages are included?                                                          \\[0.1cm]
                            & Gender                              & Are different genders included? If yes, which?                                        \\[0.1cm]
                            & Age                                 & Are different ages included? If yes, how?                                             \\[0.1cm]
                            & Accent diversity                   & Are different accents included?                                                       \\[0.1cm]
                            & Accent diversity - classes  & How many different accents are included?                                              \\[0.1cm]
                            & Accent classification              & What is the criterion to classify accents?                                            \\[0.1cm]
                            & Dialect diversity                  & Are different dialects included?                                                      \\[0.1cm]
                            & Dialect diversity - classes & How many different dialects are included?                                             \\[0.1cm]
                            & Dialect classification             & What is the criterion to classify dialects?                                           \\ \hline
 Inclusion                    & Data source                        & How were the data obtained?                                                           \\[0.1cm]
                            & Medium                              & What recording medium was used?                                                       \\[0.1cm]
                            & Dataset structure                  & Is the dataset balanced in terms of diversity?                                        \\[0.1cm]
                            & Recording Environment              & What were the conditions in the recording environment?                                \\[0.1cm]
                            & Datasheet                           & Does the dataset have a dedicated datasheet?                                          \\[0.1cm]
                            & Licence                             & Is the dataset licensed?                                                              \\[0.1cm]
                            & Licence type                       & What is the type of license?                                                          \\[0.1cm]
                            & Compensation                        & Were data-subjects compensated?                                                       \\[0.1cm]
                            & Consent                             & \begin{tabular}{@{}l@{}}Do dataset creators acknowledge that  \\they took the informed consent of data subjects?\end{tabular} \\[0.1cm] \hline
Privacy                    & Privacy                             &  \begin{tabular}{@{}l@{}}Do dataset creators took explicitly \\ into consideration the privacy of data subjects? \end{tabular}  \\[0.3cm]
                            & Privacy type                       & How do dataset creators protect the privacy of data subjects?                         \\[0.1cm] \hline
Other                       & Derivative                          & Is this dataset a derivative of another dataset?    \\ \hline                                
\end{tabular}
\label{datasetsoverview}
\end{table}

\begin{figure}
\subfloat[Linguistic diversity]{\includegraphics[width = 2.4in,valign=c]{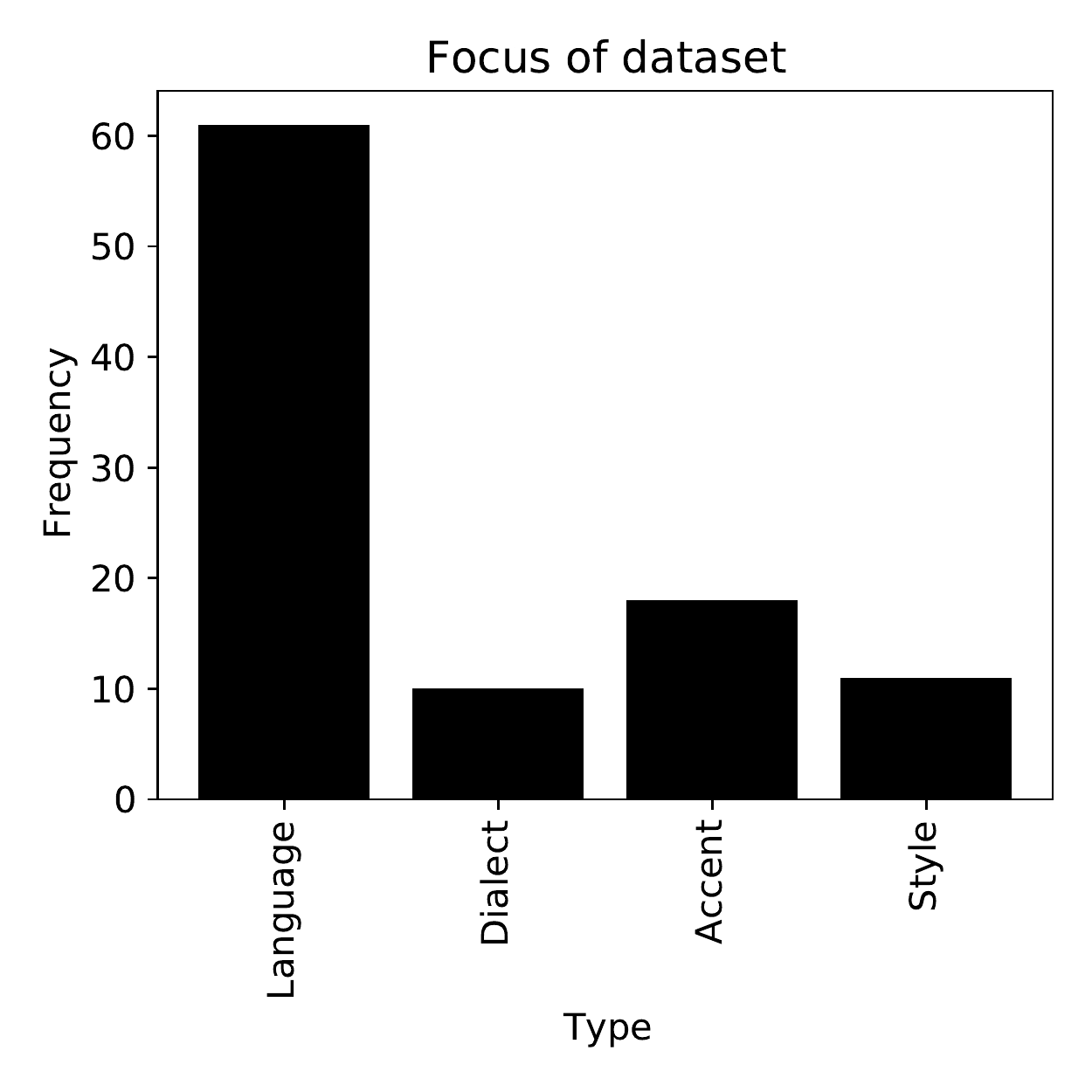}} 
\subfloat[Age diversity]{\includegraphics[width = 2.4in,valign=c]{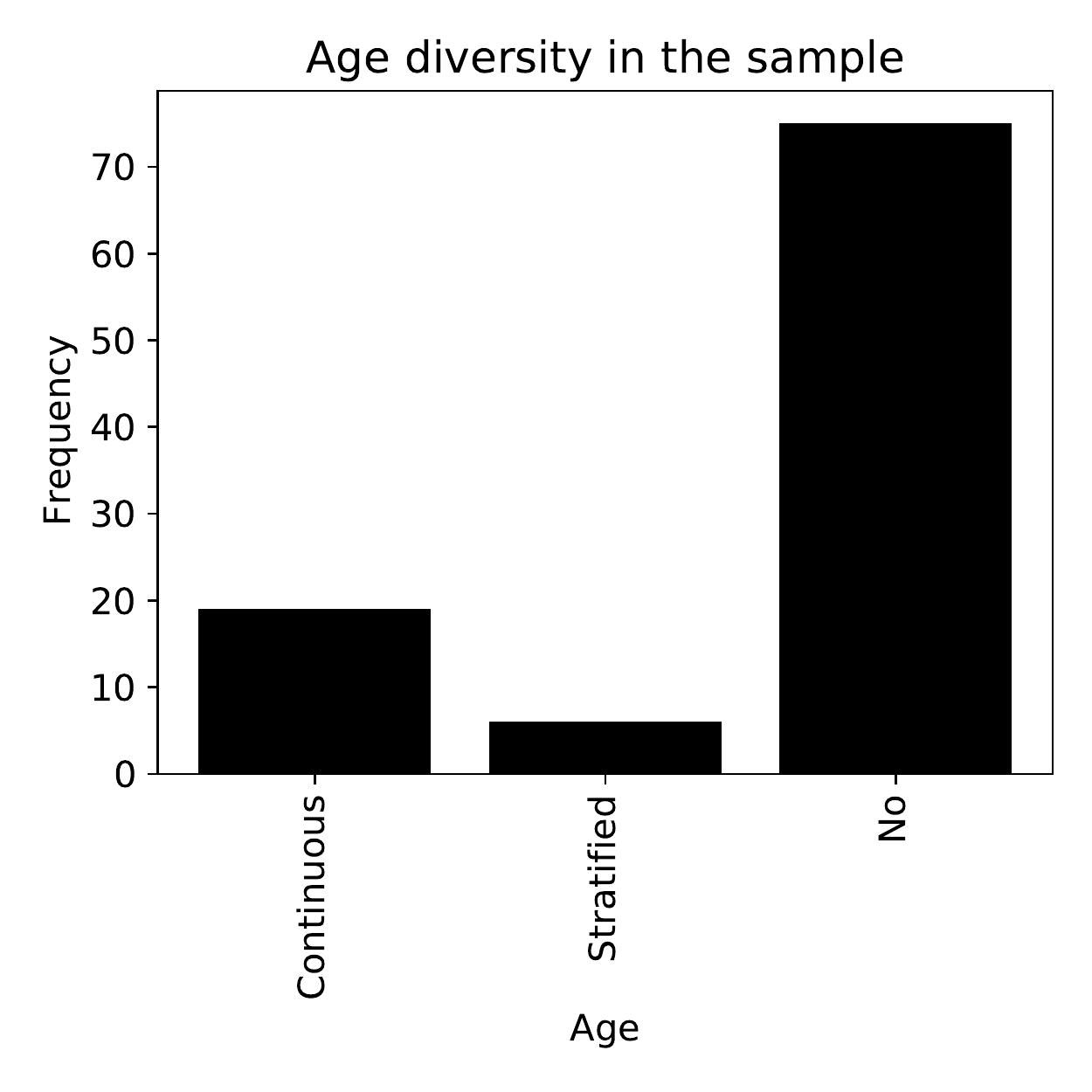}} \\
\subfloat[Gender diversity]{\includegraphics[width = 2.4in,valign=c]{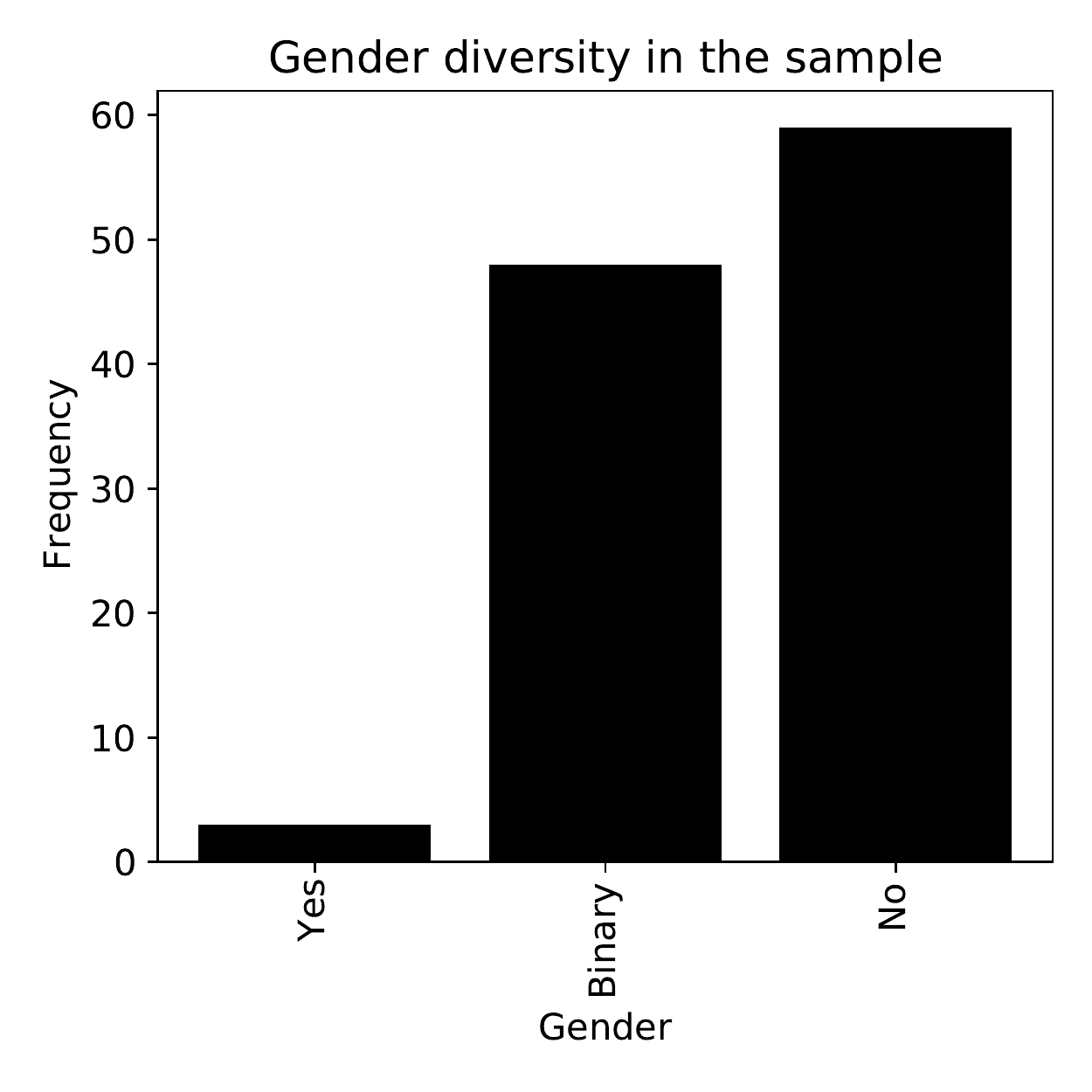}}
\subfloat[Recording environment]{\includegraphics[width = 2.4in,valign=c]{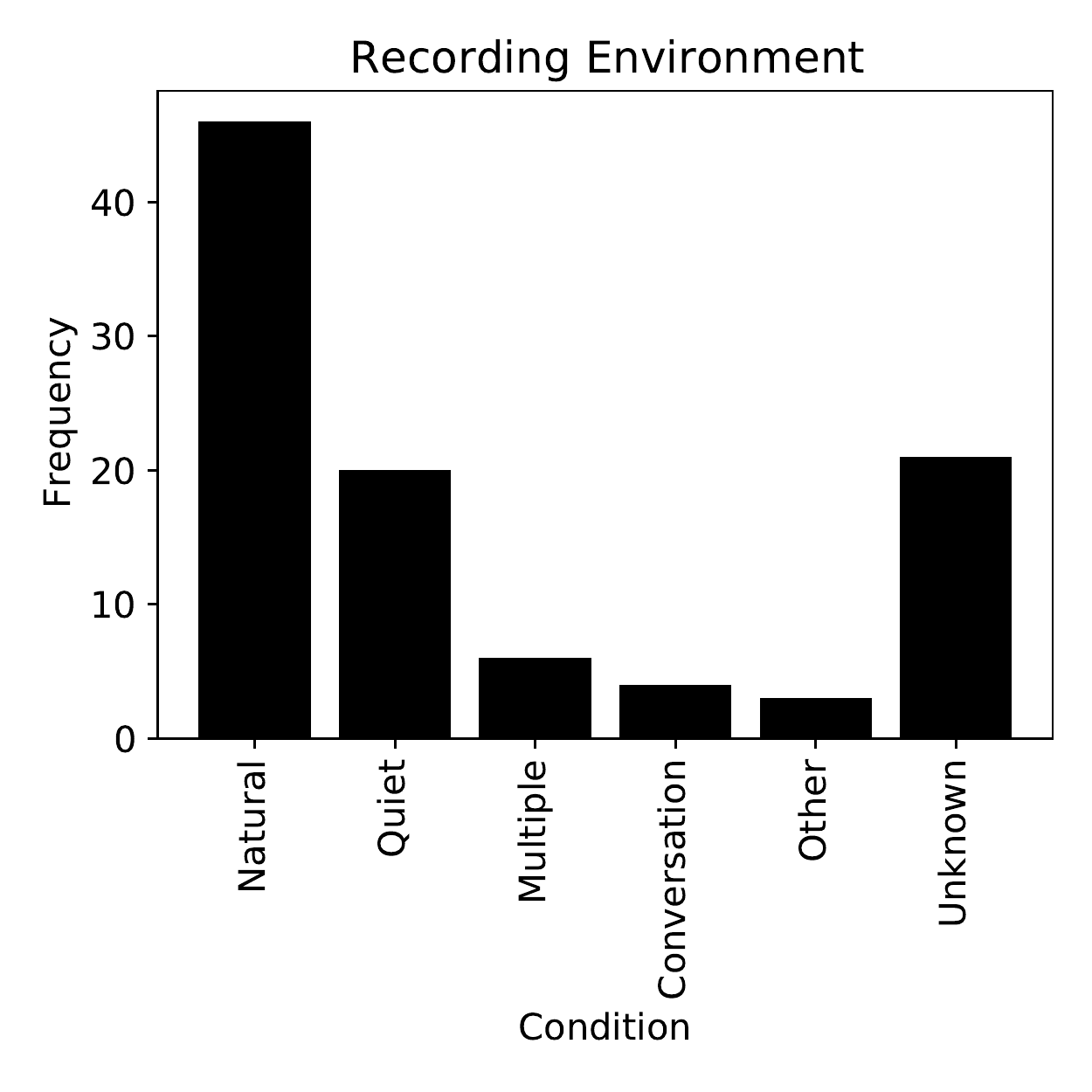}} \\
\subfloat[Recording device]{\includegraphics[width = 2.4in,valign=c]{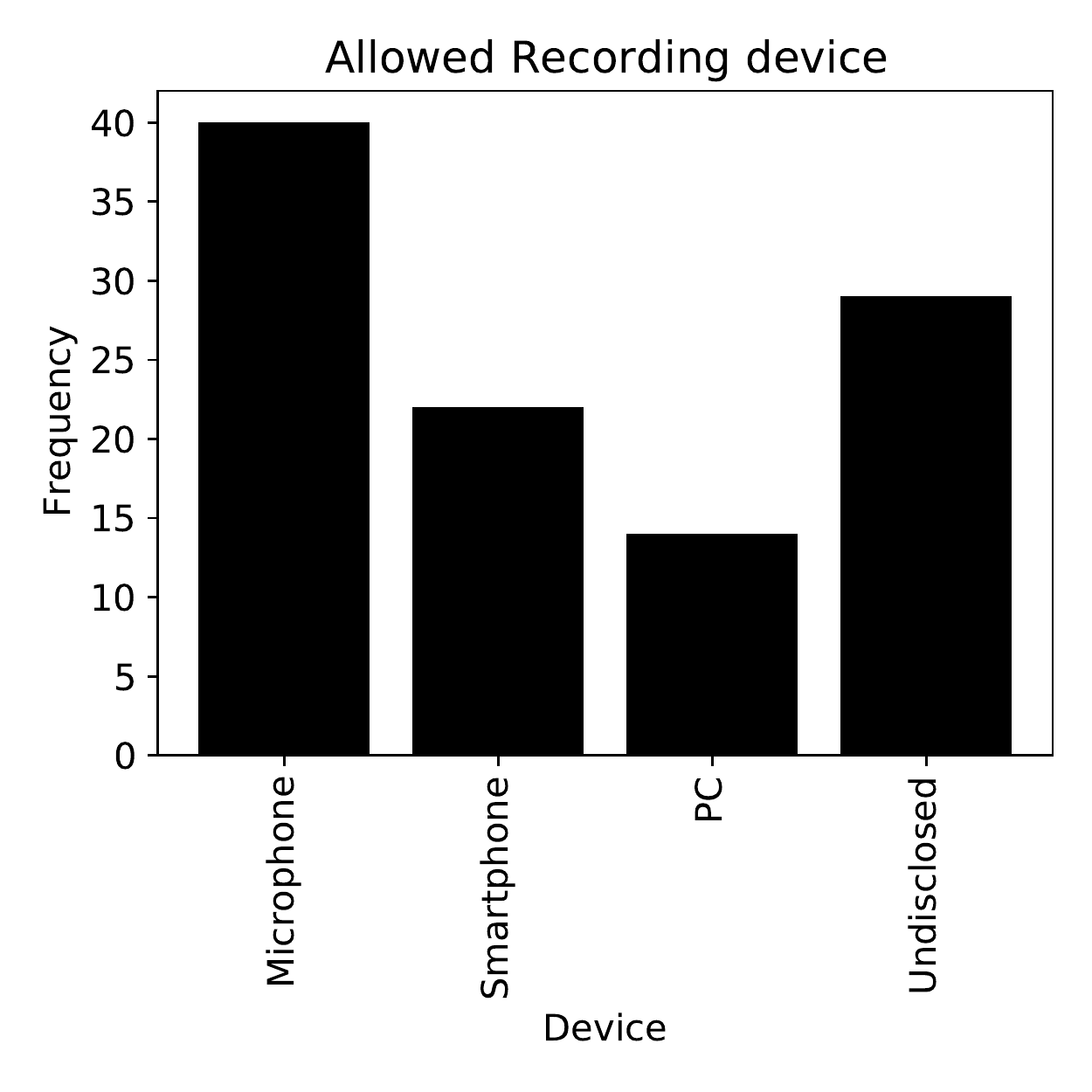}}
\subfloat[License]{\includegraphics[width = 2.4in,valign=c]{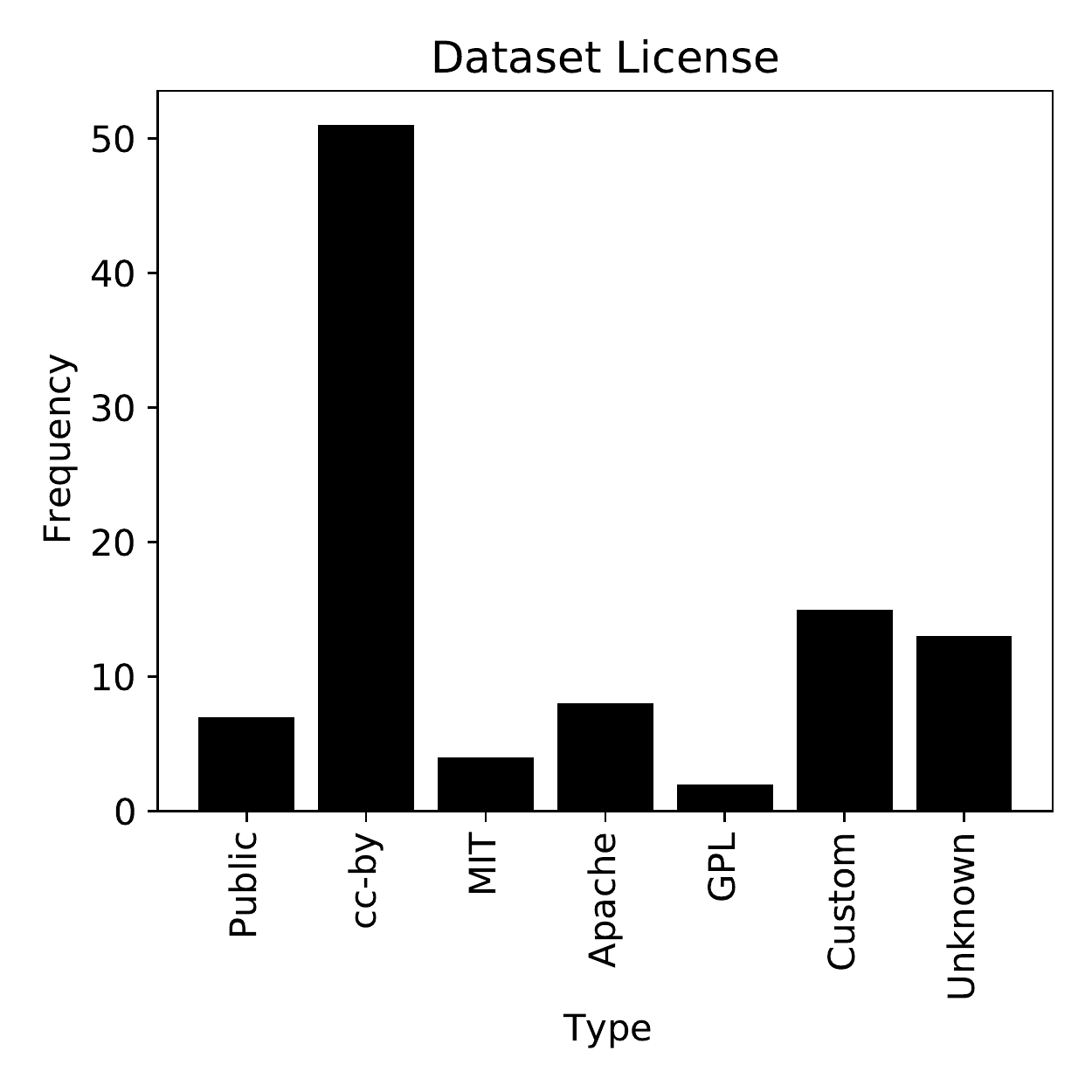}}
\caption{Descriptive statistics in the reviewed sample of speech datasets}
\label{some example}
\end{figure}
\subsection{Research studies}
\begin{figure}[htpb]
    \centering
    \includegraphics[width=1\textwidth]{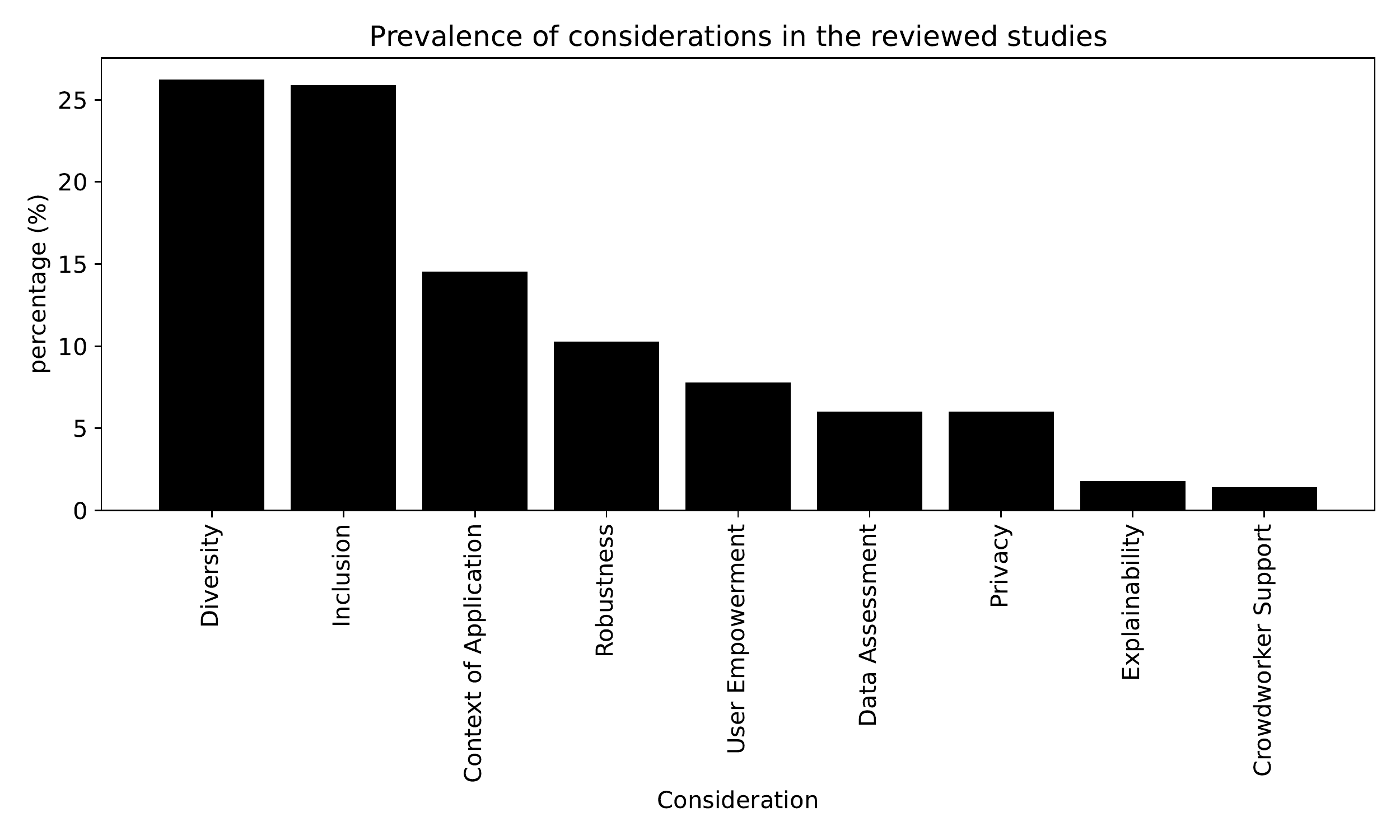}
    \caption{Distribution of ethical considerations in the reviewed sample of SLT studies (N=178). A study could contain more than one considerations.}
    \label{fig:papers}
\end{figure}

To understand issues associated with algorithmic fairness and diversity in speech, as well as to locate practices that protect and empower data-subjects and users we further collected and reviewed relevant research studies. We crawled the following research databases: ArXiv, The ACM Digital Library, Google Scholar, the ACL Anthology, and IEEE Xplore; we also collected preprints and papers published in following venues: INTERSPEECH, ICASSP, ACL conferences, ACM conferences, NeurIPS, and ICML. In total we collected 1415 papers by using the query
(Bias OR Accent OR Divers* OR Dialect OR Fairness) AND ``Speech''.  Then we manually reviewed the abstracts of the papers to ensure that they fall within the scope of our study, yielding 335 papers for in-detail exploration. We created a categorization scheme for ethical considerations related to machine learning, which can be found in Table \ref{papersoverview}. The categorization included the context of application of speech technologies, diversity \& inclusion practices, privacy \& worker protection concerns, data quality assessment, and models' robustness \& explainability techniques. Then, we excluded studies that focused on the robustness of machine learning models, since they did not have as primary focus aspects related to datasets, but rather technical and computational modeling techniques. The final list of papers we studied in detail consisted of 178 manuscripts, and can be found in the \href{https://github.com/SonyResearch/project_ethics_augmented_datasheets_for_speech_datasets}{Github repository}. An overview of the considerations in these studies can be found in Figure \ref{fig:papers}. 

\begin{table}
\caption{Overview of considerations that we matched with the sample of research studies we collected.}
\resizebox{1\textwidth}{!}{\begin{tabular}{ll}
\textbf{Category}      & \textbf{Explanation}                                                                                                                                        \\ \hline
Context of application & The study discusses issues or solutions for applying ASR systems in a specific social context.                                                             \\[0.15cm]
Diversity              & The study discusses the limitations of ASR technologies to perform well on different populations.                                                           \\[0.15cm]
Inclusion              & \begin{tabular}{@{}l@{}}The study discusses theoretic frameworks on how ASR technologies can be  \\ improved to cover more populations, dialects, accents, social groups, etc.    \end{tabular}         \\[0.3
cm]
Privacy                & The study discusses privacy issues in ASR or relates to the use of public/gray area data.                                                                    \\[0.15cm]
Robustness             & \begin{tabular}{@{}l@{}}The study discusses mathematical and technical solutions for improving the performance \\ of ASR technologies - not data collection methods/quality/properties. \end{tabular} \\[0.3cm]
User empowerment       & The study discusses ASR technologies \& dataset from a civic standpoint.                                                                                     \\[0.15cm]
Crowdworker protection & The study discusses explicitly the issues and rights of crowdworkers who create the data.                                                                    \\[0.15cm]
Data assessment        & \begin{tabular}{@{}l@{}} The study discusses explicitly how the quality of data \\ (speech segments, transcriptions) can be assessed in specific socio technical environments.     \end{tabular}      \\[0.3cm]
Explainability         & The study discusses how ASR technologies performance can be understood.     \\ \hline                                                                                 
\end{tabular}}
\label{papersoverview}

\end{table}

\section{Augmented Datasheets of Popular Datasets} \label{datareview}

We provide sample answers to the speech-specific questions posed by our augmented datasheets on five popular speech datasets: WHAM \cite{wichern2019wham}, LibriSpeech \cite{panayotov2015librispeech}, Common Voice \cite{ardila2019common}, VoxPopuli \cite{wang2021voxpopuli}, and CORAAL \cite{kendall2018corpus}. These example datasheets are available on GitHub at \url{https://github.com/SonyResearch/project_ethics_augmented_datasheets_for_speech_datasets}. Filling out the datasheets provided useful information about each of them, but also uncovered systematic gaps in their self-disclosed documentation. Focusing on questions from the motivation section of the augmented datasheets, we found that the full naming conventions of Common Voice, CORAAL, and WHAM accurately reflected the content of the datasets, but the choice of LibriSpeech and VoxPopuli were more convoluted. In terms of linguistic population identification, only some creators precisely explained why they chose specific groups to be included in the recordings (e.g., CORAAL for African American Language); others simply stated that they wanted to create a multilingual (VoxPopuli, Common Voice) or English corpus (LibriSpeech). In the disclosed information about the composition of the datasets, we were able to identify the recording hours in all of them, though the VoxPopuli creators did not explain how they chose which of the Parliament recordings to be included in the final dataset (there is a large pool of data available online that can be used), while the information about recorded hours in the Common Voice paper does not reflect the hours in the database, since the latter is constantly growing. Furthermore, we found that VoxPopuli, Common Voice and LibriSpeech had different definitions of accented speech: Common Voice had a freeform field for data subjects to self-identify; VoxPopuli classified accents based on country of origin of the speakers; and LibriSpeech used the accuracy of a standard-English ASR model to predict a segment's accent status. Regarding the content of the recordings, in no dataset did creators document whether there was speech that is highly emotional or sensitive, although this could be the case in recordings such as audio-books and parliamentary hearings. Similarly, although Common Voice is created out of Wikipedia segments, there are no details about whether the content of these sentences might have negative impact to the data-subjects due to symbolic violence. The questions about the collection process of the datasets yielded that in three datasets (VoxPopouli, Common Voice, LibriSpeech) there has been no standardization of recording media, nor an exact description about what recording medium was used for each recording. Furthermore, the VoxPopuli creators do not explain how they trained annotators that worked on creating transcriptions for a subset of the available data. Only in CORAAL were redactions specified; this was not clearly the case in the other datasets.

Our inability to answer specific datasheet questions about all datasets exemplifies the exact reason for which augmented datasheets should be necessary. Many answers require knowledge that only the dataset creators have. As all the questions listed in the datasheets are specifically built to target information that both the creators and the users should have knowledge of in order to utilize the dataset in a fair and precise way, it is crucial for creators to take the datasheet as their guide for revision and incorporate the missing information.  Taking the above into consideration, the application of new questions in the augmented datasheets showed that there is a considerable amount of improvement dataset creators can make in transparently informing users about the properties of their dataset. Most importantly, when using these questions as a foundation for developing a dataset, they can make more ethically grounded design choices.  
\newpage
\section{Examples of Transcription} \label{software}

\begin{figure}[h]
    \centering
    \includegraphics[width=0.6\textwidth]{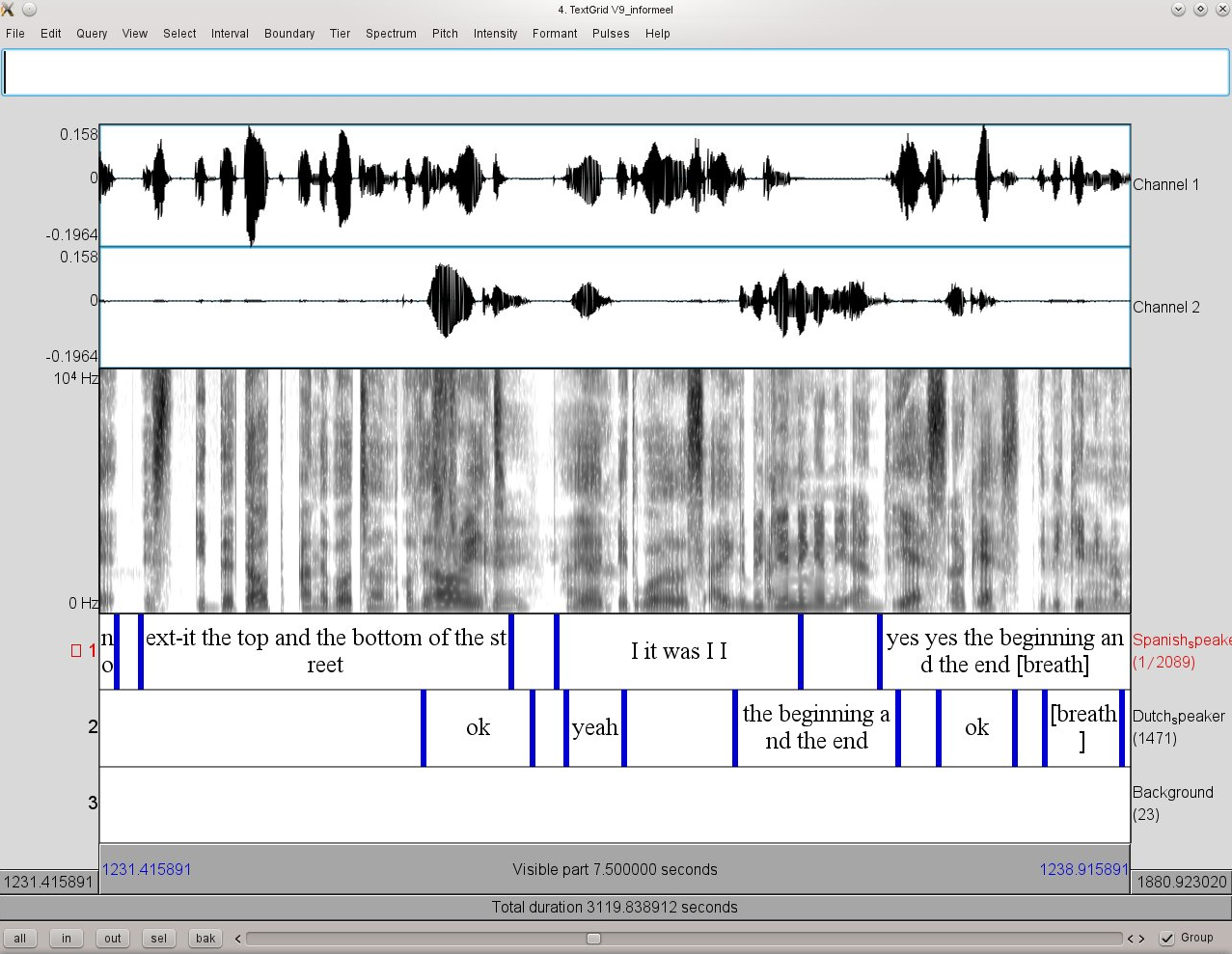}
    \caption{Transcription made in Praat, example taken from \cite{kouwenhoven2018register}. Because the data was collected through Dutch-speaking confederates, the tiers are named `Spanish speaker' and `Dutch speaker'. The tiers include the orthographic transcription of each speakers' speech, broken into segments accordingly, as well as any human-made noise defined prior, such as [breath] or [laugh]. There is a separate tier for non-human-made noise, named `Background'.}
    \label{fig:spanish_dutch}
\end{figure}

\begin{figure}[h]
    \centering
    \includegraphics[width=0.6\textwidth]{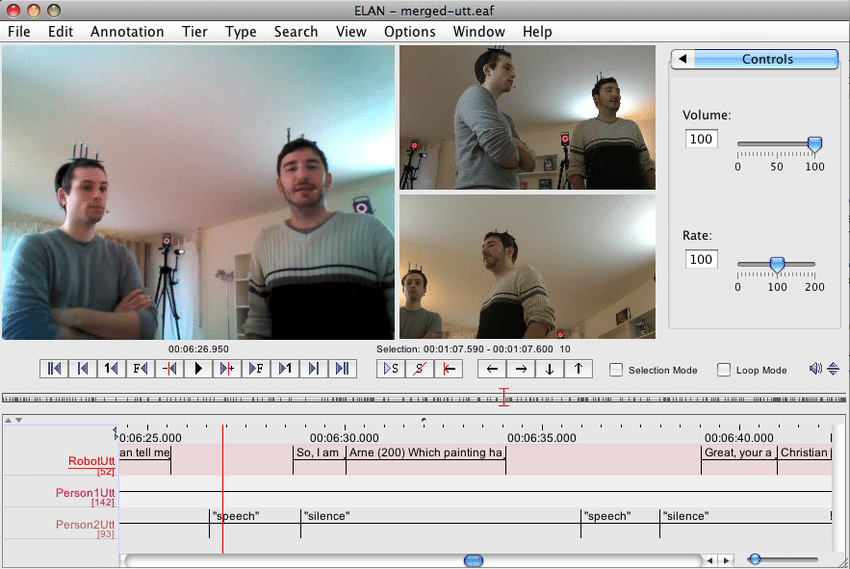}
    \caption{Transcription made in ELAN, example taken from \cite{jayagopi2012vernissage}. The dataset annotated not only speech but head location and movement of the speakers, thus needing both audio and visual tiers. Because the data was collected by German researchers, the audio tier specifically marked where the speech was produced in German. Instead of detailing different types of noise, the dataset categorized the utterances into three types: \textit{speech}, \textit{silence} and \textit{laughter}.}
    \label{fig:elan}
\end{figure}

\section{Augmented Datasheets for Speech Datasets TEMPLATE} 
\includepdf[pages=-]{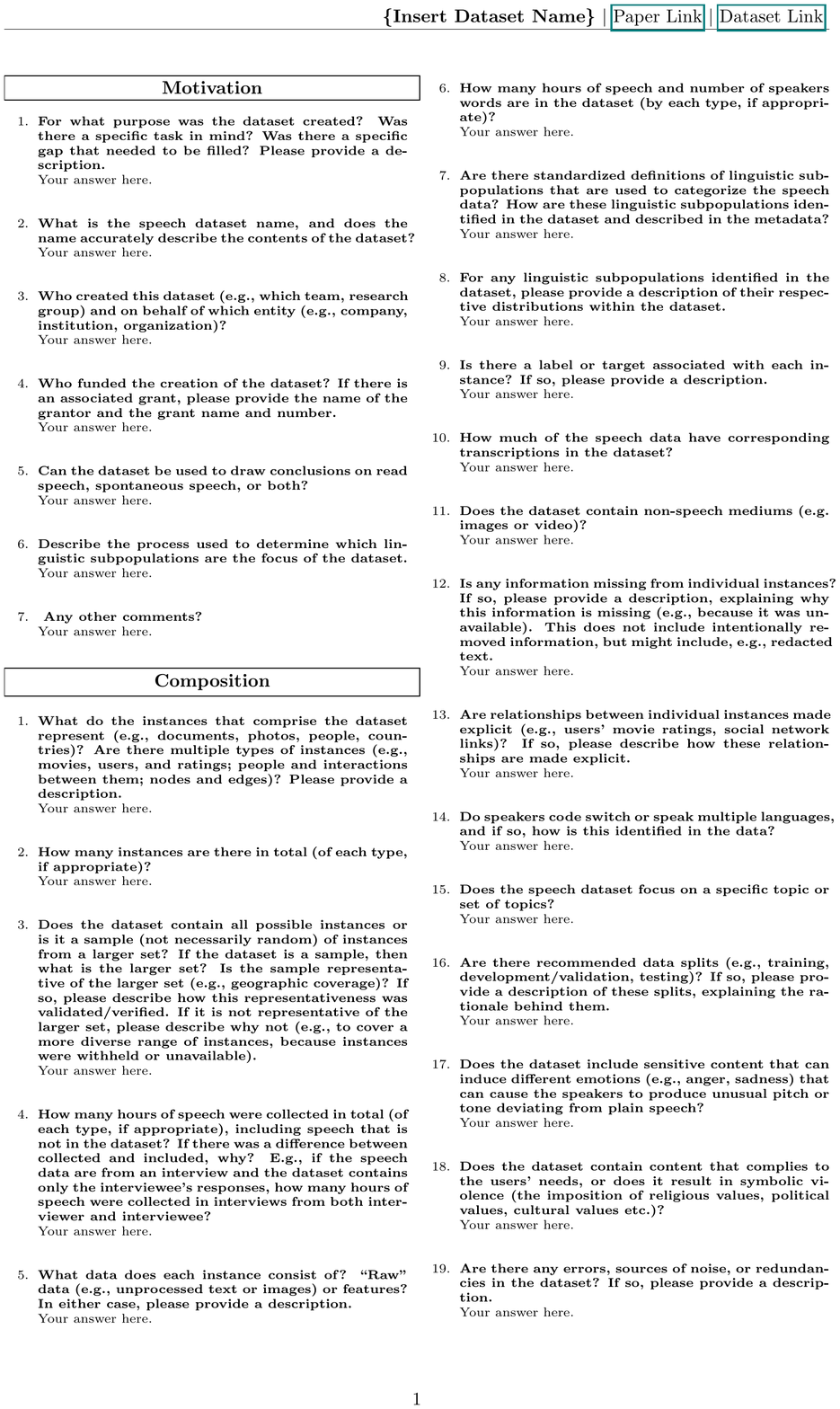}

\end{document}